\documentclass[iop]{emulateapj}
\usepackage{epsfig}
\usepackage{graphicx}
\usepackage{longtable}
\usepackage{natbib}
\usepackage{subfigure}
\usepackage{amsmath}
\usepackage{xcolor}
\newcommand{\Msolar}{M$_{\odot}$}
\newcommand{\Lsolar}{L$_{\odot}$}
\newcommand{\Rsolar}{R$_{\odot}$}

\newcommand{\PRV}{P$_{RV}$}
\newcommand{\PPM}{P$_{\mu}$}
\newcommand{\pms}{$\pm$}

\newcommand{\ergss}{ergs s$^{-1}$}

\shorttitle{On The Origin of Sub-subgiants.}

\shortauthors{Leiner et al.}

\begin{document}

\title{On the Origin of Sub-Subgiant Stars II: Binary Mass Transfer, Envelope Stripping, and Magnetic Activity \footnote{This is paper number 74 in the WIYN Open Cluster Study}}

\author{Emily Leiner\altaffilmark{1}, Robert
  D. Mathieu\altaffilmark{1}, and Aaron
  M. Geller\altaffilmark{2,}\altaffilmark{3} }  
\email{leiner@astro.wisc.edu}

\altaffiltext{1}{Department of Astronomy, University of Wisconsin-Madison, 475 North Charter St, Madison, WI 53706, USA}
\altaffiltext{2}{Center for Interdisciplinary Exploration and Research in Astrophysics (CIERA) and Department of Physics and Astronomy, Northwestern University, 2145 Sheridan Rd, Evanston, IL 60208, USA}
\altaffiltext{3 }{Adler Planetarium, Department of Astronomy, 1300 S. Lake Shore Drive, Chicago, IL 60605}

\begin{abstract}

Sub-subgiant stars (SSGs) lie to the red of the
main-sequence and fainter than the red giant branch in cluster color-magnitude diagrams (CMDs), a region not easily
populated by standard stellar evolution pathways. While there has
been speculation on what mechanisms may create these unusual stars, no
well-developed theory exists to explain their origins. Here we
discuss three hypotheses of SSG formation: (1) mass transfer in a
binary system, (2) stripping of a subgiant's envelope, perhaps during a dynamical encounter, and (3) reduced luminosity due to magnetic fields that lower
convective efficiency and produce large star spots. Using the stellar evolution code MESA, we develop evolutionary tracks for each of these hypotheses, and compare the expected stellar and orbital properties of these models with six known SSGs in the two open clusters M67 and NGC 6791. All three of these mechanisms can create stars or binary systems in the SSG CMD domain.  We also calculate the frequency
with which each of these mechanisms may create SSG systems, and find that the magnetic
field hypothesis is expected to create SSGs with the highest frequency in open clusters. Mass transfer and envelope stripping have lower expected formation frequencies, but may nevertheless create occasional SSGs in open clusters. They may also be important mechanisms to create SSGs in higher mass globular clusters. \
\end{abstract}


\section{Introduction}

Optical color-magnitude diagrams reveal that $25\%$ of the evolved stars in
older open clusters do not fall along standard single-star evolutionary
tracks. These stars include the well-known blue stragglers, but also the 
yellow giants and sub-subgiants. 

Sub-subgiant stars (SSGs) were first identified in the color-magnitude diagram
(CMD) of the open cluster M67 \citep{Belloni1998, Mathieu2003}. These two SSGs fall to the red of both the main-sequence and main-sequence binary track and well below the subgiant branch. Both SSGs have high membership probabilities based on both proper-motion and radial-velocity (RV) data, leaving a negligible probability that both are field interlopers. 

Broadly speaking, the populations of SSGs in
globular and open clusters share similar characteristics. They fall to the red of the main sequence and below the subgiant and giant branch on optical CMDs, a region that can not be easily populated by either
single-star evolutionary theory or by any combination of two normal cluster
stars. They are also typically X-ray sources with $L_x \sim 10^{30}- 10^{31}$
erg s$^{-1}$ and photometric variables with periods between 1 and 20
days. Where binary status is known, they are often found to be close binary systems with orbital periods on the order of 1-10 days. Similar X-ray sources and photometric variables are also found to red of the RGB. We call these stars ``red stragglers" rather than sub-subgiants, though the two types may be related and have similar formation mechanisms. Geller et al. 2017a give a census of the open cluster and globular cluster red stragglers and SSGs known from the literature. 

No well-developed theory has yet been presented for the origin and
evolutionary status of these non-standard stars. \citet{Mathieu2003} suggest
they may be products of close stellar encounters involving binaries, or stars with enhanced extinction (i.e.~due to the presence of
circumstellar material). Other authors invoke mass transfer and stellar
collision events to form SSGs \citep{Hurley2005, Albrow2001}. 

While many SSGs are kinematic cluster members, most do not have binary orbital information (Geller et al. 2017a). The sample of SSGs with both high quality 3D kinematic memberships and known orbital solutions for the binaries is small, consisting of 6 stars in two open clusters in the WIYN Open Cluster Study (WOCS; \citealt{Mathieu2000}): 4 SSGs (3 binaries and one single star) in NGC 6791 \citep{Platais2011, Milliman2016} and 2 binaries in M67 \citep{Mathieu2003}. We use this
sample to guide the formation of an origin theory that matches the
observed properties of this sample. We focus on three hypotheses for SSG formation: mass transfer in a binary system,
stripping of a subgiant's envelope, 
or a reduced luminosity due to the presence of a strong magnetic field. 

\section{Sub-subgiant Sample and Observations}
\begin{figure*}[htbp]
\subfigure{\includegraphics[width=.49\linewidth ]{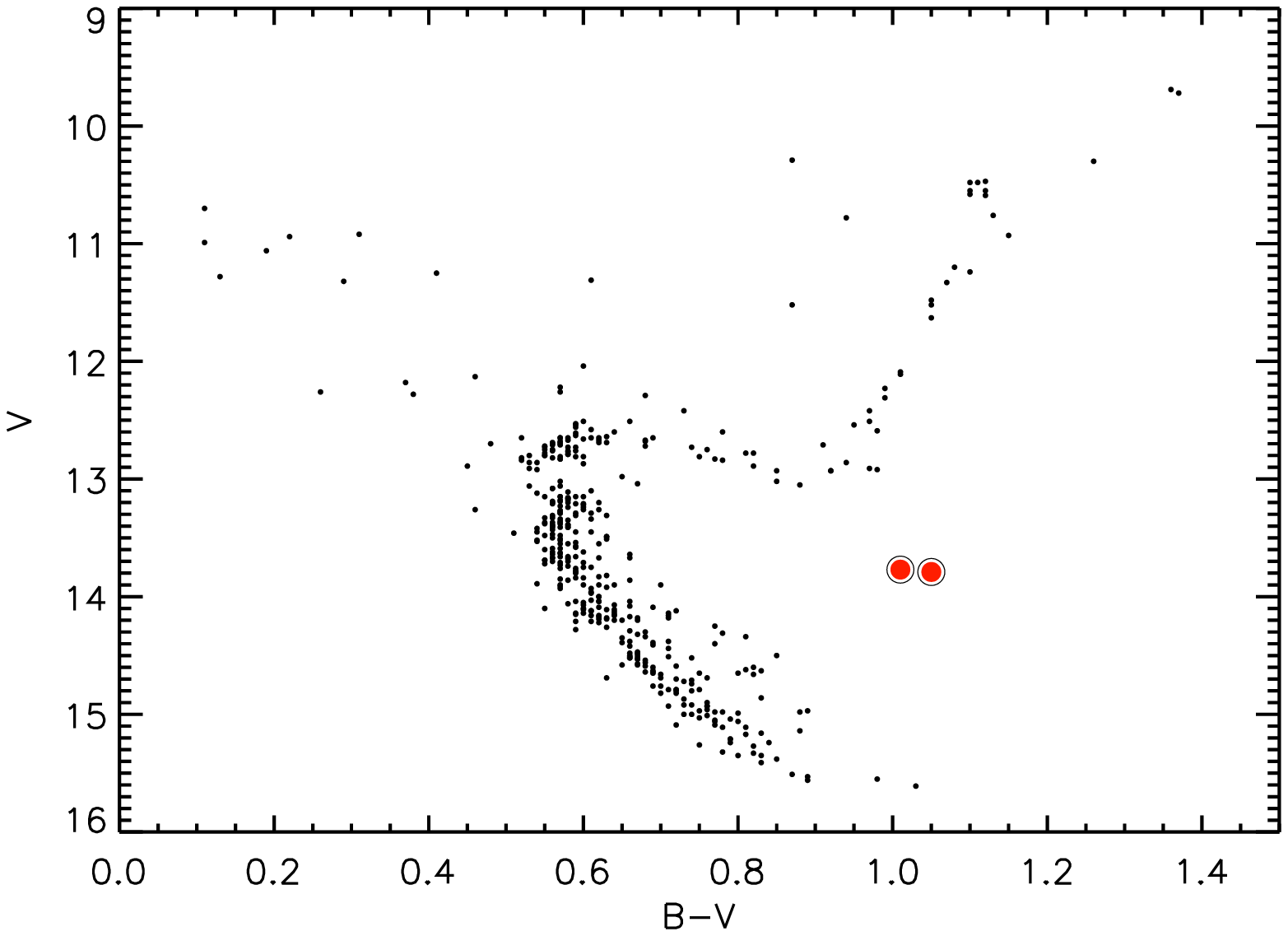}}
\subfigure{\includegraphics[width=.49\linewidth ]{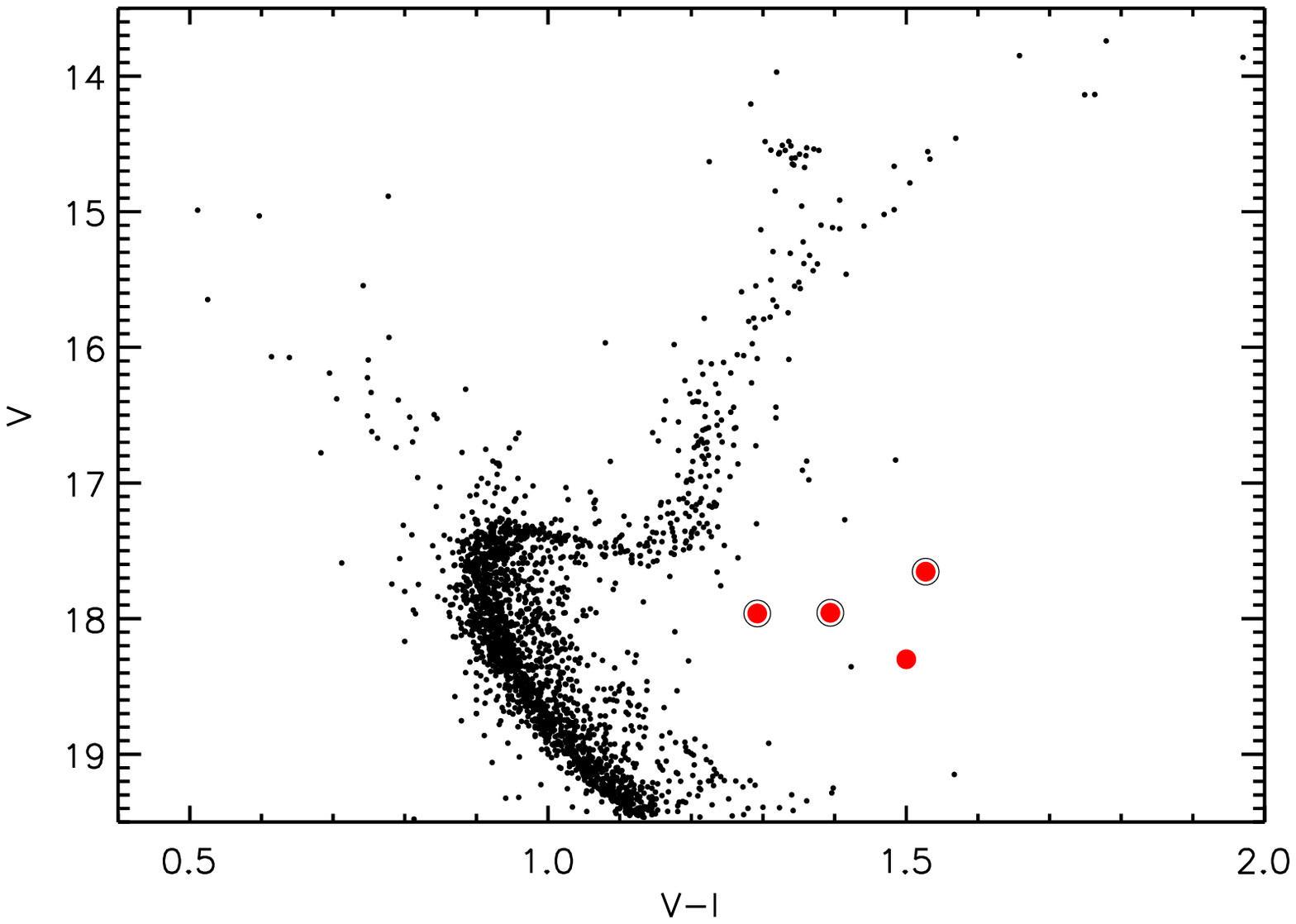}}
\caption{\textbf{(left)} A BV color-magnitude diagram of M67 showing all 3D kinematic members \citep{Geller2015}. The sub-subgiants are shown with red circles. \textbf{(right)} A VI CMD showing proper-motion members of NGC 6791 \citep{Platais2011}. The SSGs confirmed to be 3D kinematic
members of NGC 6791 are shown in red circles \citep{Milliman2016}. The binary SSGs in both plots are circled in black. \label{cmds}}
\end{figure*}
 
\subsection{M67 and NGC 6791 Cluster Properties}
Our sample of six SSGs is drawn from two WOCS open clusters: M67 and NGC 6791. CMDs for both clusters are shown in Figure~\ref{cmds} with the locations of the SSGs highlighted. 

Located at $\alpha=8^h51^m23^s.3$, $\delta=+11^{\circ}49^{\arcmin}02^{\arcsec}$ (J2000), M67 is an old, solar-metallicity open cluster (e.g. \citealt{Montgomery1993, Taylor2007}). Distance measurements for the cluster range from 800-900 pc, with reddening measurements ranging from E$(B-V)=0.015$ to 0.056 \citep{Geller2015}. For this study, we adopt E$(B-V)=.041$ \citep{Taylor2007} and $(m-M)_{0}=9.7$ \citep{Sarajedini2009}. Age determinations put the cluster at around 4 Gyr (e.g. \citealt{Montgomery1993, vandenBerg2004}) with a main-sequence-turnoff mass of $\sim1.3$ \Msolar. 

Located at $\alpha=19^h20^m58^s.09$, $\delta=+37^{\circ}46^{\arcmin}31^{\arcsec}$ (J2000), NGC 6791 is an old (8 Gyr; \citealt{Carney2005, Grundahl2008}) and metal-rich ([Fe/H]=+0.40, \citealt{Carney2005}) open cluster. Distance measurements put the cluster at around 4 kpc (e.g. \citealt{Grundahl2008}). The turn-off mass of the cluster is $\sim1.1$ \Msolar~\citep{Brogaard2012}. 
 For this study we use the distance modulus and reddening values found by \citet{Carney2005}: E$(B-V)=0.14$, $(m-M)_0=13.07$.

\subsection{SSG Cluster Memberships and Orbital Parameters}
In Table 1 we list the WOCS ID, coordinates, proper-motion membership probabilities (\PPM), and the radial-velocity membership
probabilities (\PRV) for the six SSGs in our sample. We also include the $BVI$ photometry from
\cite{Stetson2003} for NGC 6791, and \citet{Montgomery1993} for M67. In the comments section we include other
identifiers for these targets from previous studies. 

Five of the six SSGs in our sample are binary systems, and for these we also list periods and eccentricities in Table 1. One SSG is a double-lined spectroscopic binary (SB2; WOCS 15028), and the other four are single-lined (SB1s). All but one of these binaries are circular, and they all have short periods ranging from 2.8 to 18.4 days. 

While it is possible that  any one SSG could be a
field contaminant, given the kinematic memberships  the probability that all of these systems are field stars
is quite low. \citet{Mathieu2003} provide a membership analysis for the M67 SSGs, calculating a $9\%$ probability that one of the 246 3D kinematic members in their sample is a nonmember. The probability of  finding 2 nonmembers is just 0.4$\%$. This is within their entire sample of kinematic members, so the
likelihood that the 2 SSGs specifically are field stars is smaller still. 

\citet{Milliman2016} provide a similar analysis of the NGC 6791
SSGs. Their analysis, based on the kinematic membership probabilities and the CMD location of the stars, indicates that it is highly unlikely for all four stars to be field contaminants. Specifically, they calculate a 17\% probability that one of the four SSGs is a field star, dropping to just 1.8\% chance that two SSGs are field stars, 0.13\% for 3, and 0.007 \% for all 4. We are thus confident that our sample of 6 SSGs cannot be explained simply by field contamination.

\begin{deluxetable*}{cccccccccccrrc}[htbp]
\tablewidth{0pt}
\centering
\tabletypesize{\footnotesize}
\tablecaption{M67 and NGC 6791 SSGs\label{tab:targets}}
\tablehead{ \colhead{Cluster} & \colhead{WOCS ID} & \colhead{$\alpha$ (J2000)}
& \colhead{$\delta$ (J2000)} &  \colhead{$P_\mathrm{\mu}\tablenotemark{a}$(\%)} & \colhead{\PRV}\tablenotemark{b}&
  \colhead{$V$} & \colhead{$B-V$} & \colhead{$V-I$} &  \colhead{$P_\text{orb}$\tablenotemark{c}} (days)& \colhead{e\tablenotemark{c}} & \colhead{Other IDs\tablenotemark{d}} }
\startdata
   M67      & 15028  & 08 51 25.30 & +12 02 56.3  & 97 &99 &13.77 & 1.01    &\nodata& 2.823094  & 0 & S1113\\  
   & & & & & & & & &$ \pm 0.000014$ & $\pm 0$ & \\
   \\
  M67      & 13008  & 08 51 13.36 & +11 51 40.1  & 98 &98 &13.79 & 1.05    &\nodata& 18.396 & 0.26 & S1063\\
     & & & & & & & & &$ \pm 0.005$ & $\pm 0.014$& \\
     \\
   NGC 6791 & 130013 & 19 21 25.22 & +37 45 49.82 & 99 &84 &17.65 & \nodata & 1.53 & 7.7812  & 0.015&  15561 \\
      & & & & & & & & & $\pm 0.0012$ &$ \pm 0.019$ & \\
      \\
   NGC 6791 & 131020 & 19 20 10.61 & +37 51 11.20 & 96 &85 &18.30 & \nodata & 1.50 & \nodata & \nodata &  83 \\
   & & & & & & & & & \nodata & \nodata & \\
   \\
   NGC 6791 & 147014 & 19 20 21.48 & +37 48 21.60 & 99 &95 &17.96 & 1.35    & 1.39 & 11.415& 0.05 &   746  \\
      & & & & & & & & & $\pm 0.007$ & $\pm 0.04$ & \\
      \\
   NGC 6791 & 170008 & 19 20 38.88 & +37 49 04.29 & 99 &63 &17.96 & 1.15    & 1.29 & 5.8248  & 0.013 & 3626\\
         & & & & & & & & & $\pm 0.0008 $ & $\pm 0.020$ & \\
\enddata
\tablenotetext{a}{Proper motion probabilities come from \citet{Girard1989} for
M67 and from \citet{Platais2011} for NGC 6791}
\tablenotetext{b}{RV membership probability from \citet{Geller2015} for M67
 and \citet{Milliman2016} for NGC 6791}
 \tablenotetext{c}{Periods (P$_\text{orb}$) and eccentricities ($e$) are taken from \citet{Milliman2016} for NGC 6791 and \citet{Mathieu2003} for M67}
\tablenotetext{d}{Comments list \citet{Stetson2003} IDs for NGC 6791 and Sanders IDs
  (proceeded by an S) for M67}
\end{deluxetable*}

\subsection{SSG Spectral Energy Distributions}
In order to measure the physical characteristics of the open cluster SSGs, we
pieced together spectral energy distributions (SEDs) from existing optical
observations \citep{Montgomery1993, Stetson2003} and photometry from the Two-Micron All-Sky
Survey (2MASS; \citealt{2MASS}), Wide Field Infrared Explorer (WISE;
\citealt{WISE}), and Spitzer Space Telescope Infrared Array Camera \citep{Skrutskie2007}. We used these SEDs
to fit a temperature and radius to each star and determine the bolometric luminosities of the systems.

\begin{deluxetable*}{l r r r r r r r r r r r }[htbp]
\centering
\tabletypesize{\scriptsize}
\tablewidth{0pt}
\tabletypesize{\footnotesize}
\tablecaption{Photometry for SSGs\tablenotemark{a}\label{tab:photometry}}
\startdata
\tablehead{\colhead{WOCS ID} & \colhead{U} &\colhead{B}
 &\colhead{V} &\colhead{R} &\colhead{I} &\colhead{J} &\colhead{H}
  &\colhead{K} &\colhead{W1} &\colhead{W2} &\colhead{W3} }
  15028\tablenotemark{b}  &    15.3 &    14.78   &   13.77 &    13.09 &  \nodata &   11.671   &  11.123  &   10.971  &10.84  &  10.822   &  10.681 \\
& \nodata &\nodata&\nodata&\nodata&\nodata & \pms.021&\pms.023&\pms.022&\pms.023&\pms.021&\pms.086\\
\\
   13008\tablenotemark{b}   & 15.56 &   14.84 &   13.79  & \nodata &12.59   &  11.657    &  11.058   & 10.958 &10.810   &  10.855 &10.643 \\
&\nodata&\nodata&\nodata&\nodata&\nodata&\pms.022&\pms.019&\pms.018&\pms.022&\pms.019&\pms.101\\
\\
130013\tablenotemark{c}  & \nodata &    \nodata   &  17.654  & \nodata &  16.127 & 15.197   &  14.495  &  14.485 &   14.16 &   14.337 \\
       & \nodata &    \nodata &  \pms .0052& \nodata& \pms .0117& \pms.047&\pms.050&\pms.086&\pms.028&\pms.043&\nodata\\
       \\
131020\tablenotemark{c}  &\nodata &\nodata& 18.3&\nodata & 16.8 & 15.791 & 15.040 & 14.780 & \nodata& \nodata& \nodata\\
&\nodata& \nodata &\nodata&\nodata&\nodata&\pms.069&\pms.071&\pms.101&\nodata&\nodata&\nodata\\
\\

147014\tablenotemark{c}   &  \nodata &   19.305 &  17.957  & \nodata &     16.563  &  15.532  &  14.810  &  14.707  &  14.643  &  14.77 &   \nodata  \\
      &  \nodata & \pms.0076& \pms .0028& \nodata&\pms .01 & \pms.059&\pms.057&\pms.101&\pms.033&\pms.051&\nodata\\
      \\
 170008\tablenotemark{c}  & \nodata&  19.116  &  17.962   & \nodata &    16.670 &  15.795  &   15.248  & 15.055  &  \nodata & \nodata&\nodata  \\
  & \nodata&\pms .0069&\pms .0012&\nodata&\pms.0011&\pms.061&\pms.08&\pms.1029&\nodata&\nodata&\nodata\\

\enddata

\tablenotetext{a}{Errors listed are measurement errors, not variability.}
\tablenotetext{b}{UBVR photometry from \citet{Montgomery1993}.}
\tablenotetext{c}{BVI photometry from \citet{Stetson2003}.}

\end{deluxetable*}

\subsubsection{SED Fitting}
\label{sec:SEDfits}
\begin{figure*}
\begin{center}
\subfigure{\includegraphics[width=.32\linewidth ]{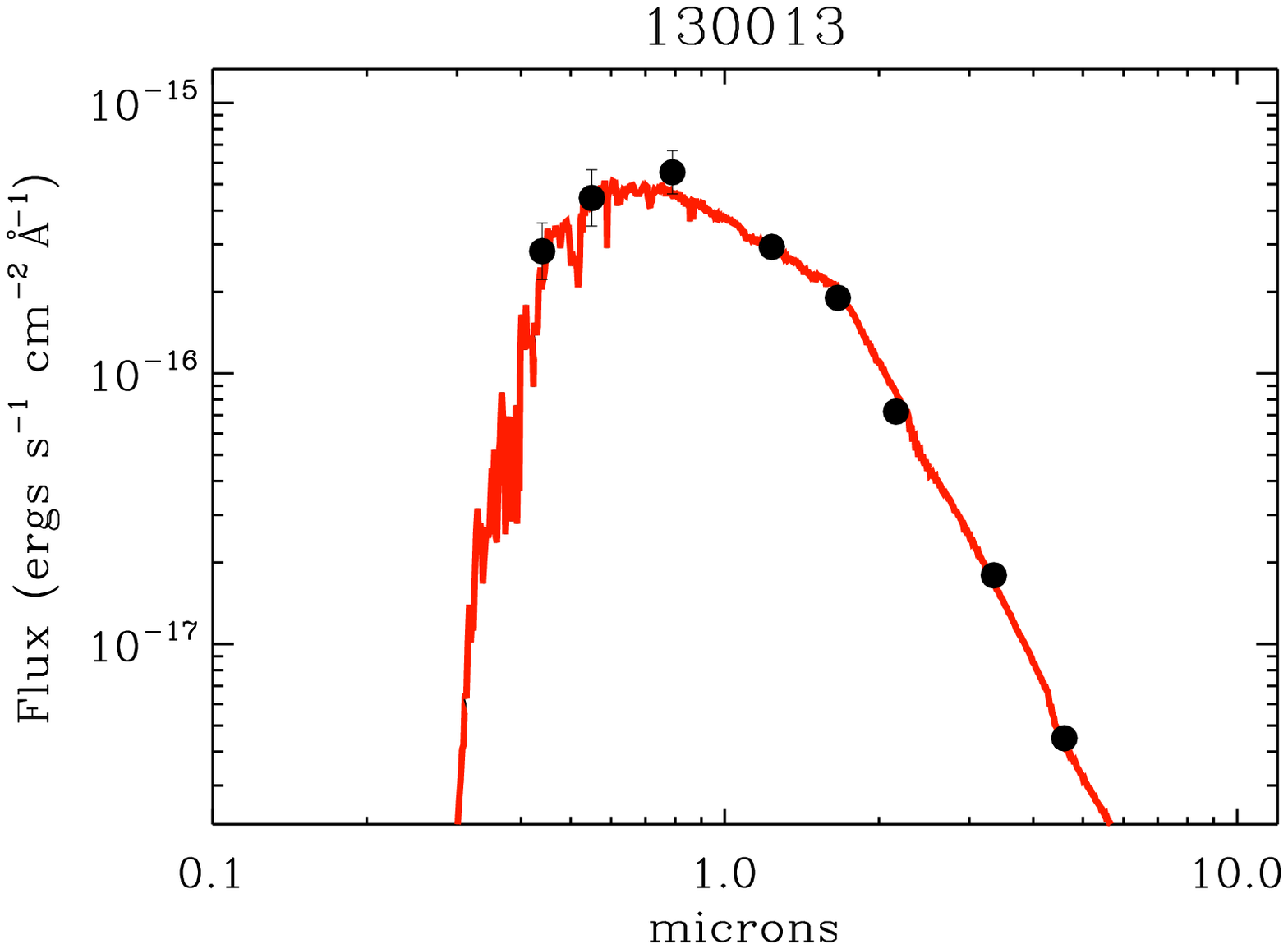}}
\subfigure{\includegraphics[width=.32\linewidth ]{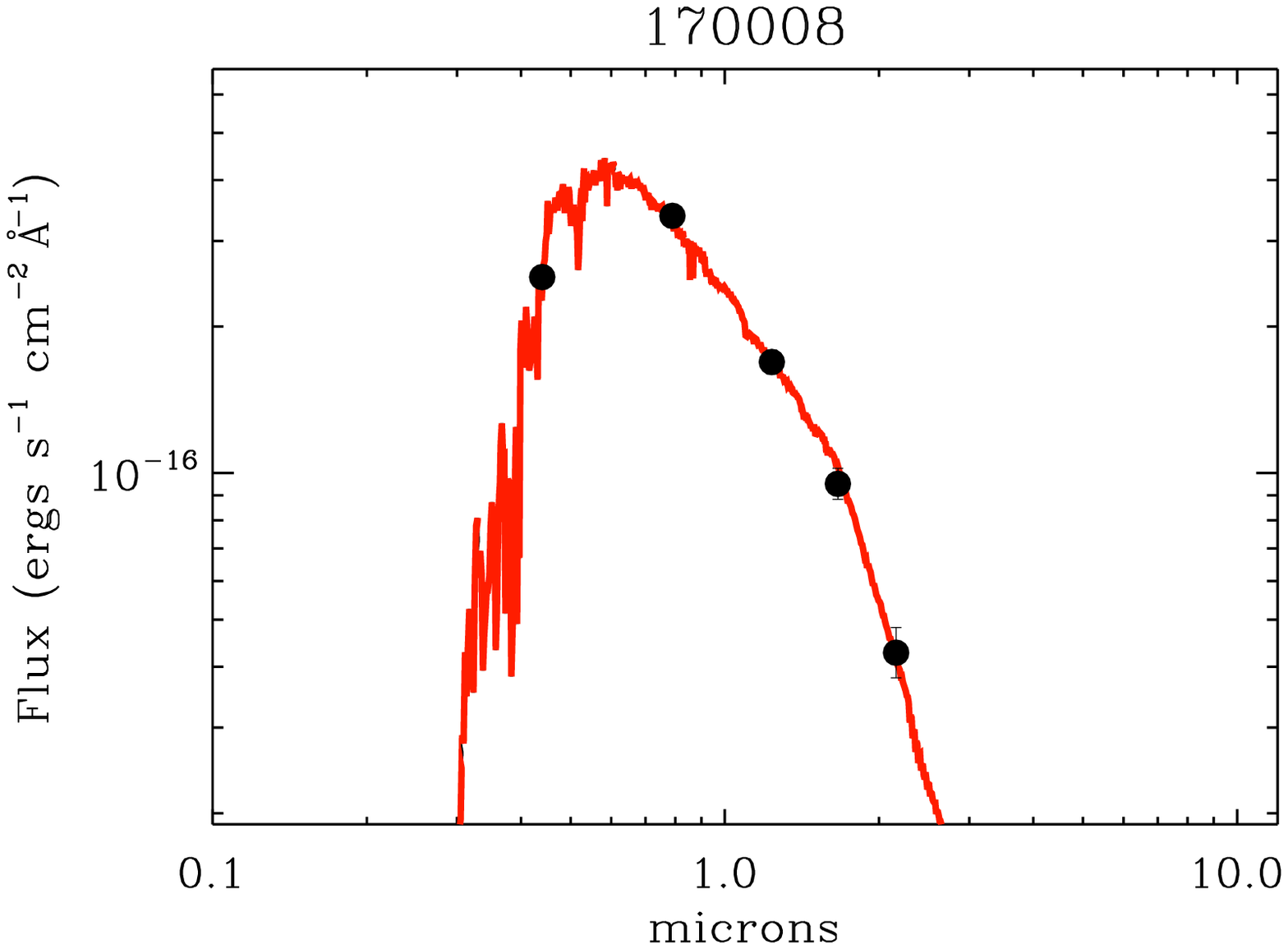}}
\subfigure{\includegraphics[width=.32\linewidth ]{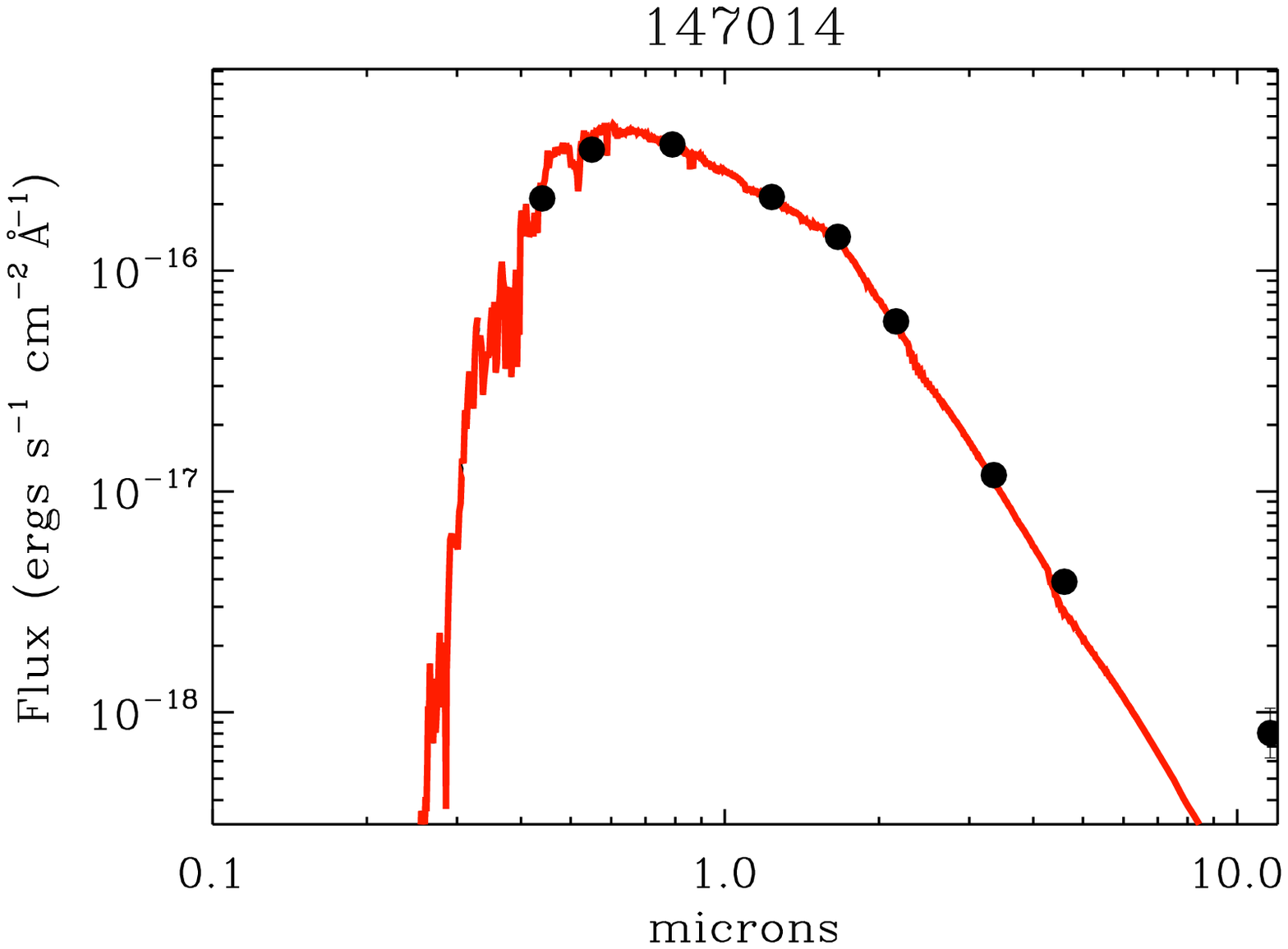}}
\subfigure{\includegraphics[width=.32\linewidth ]{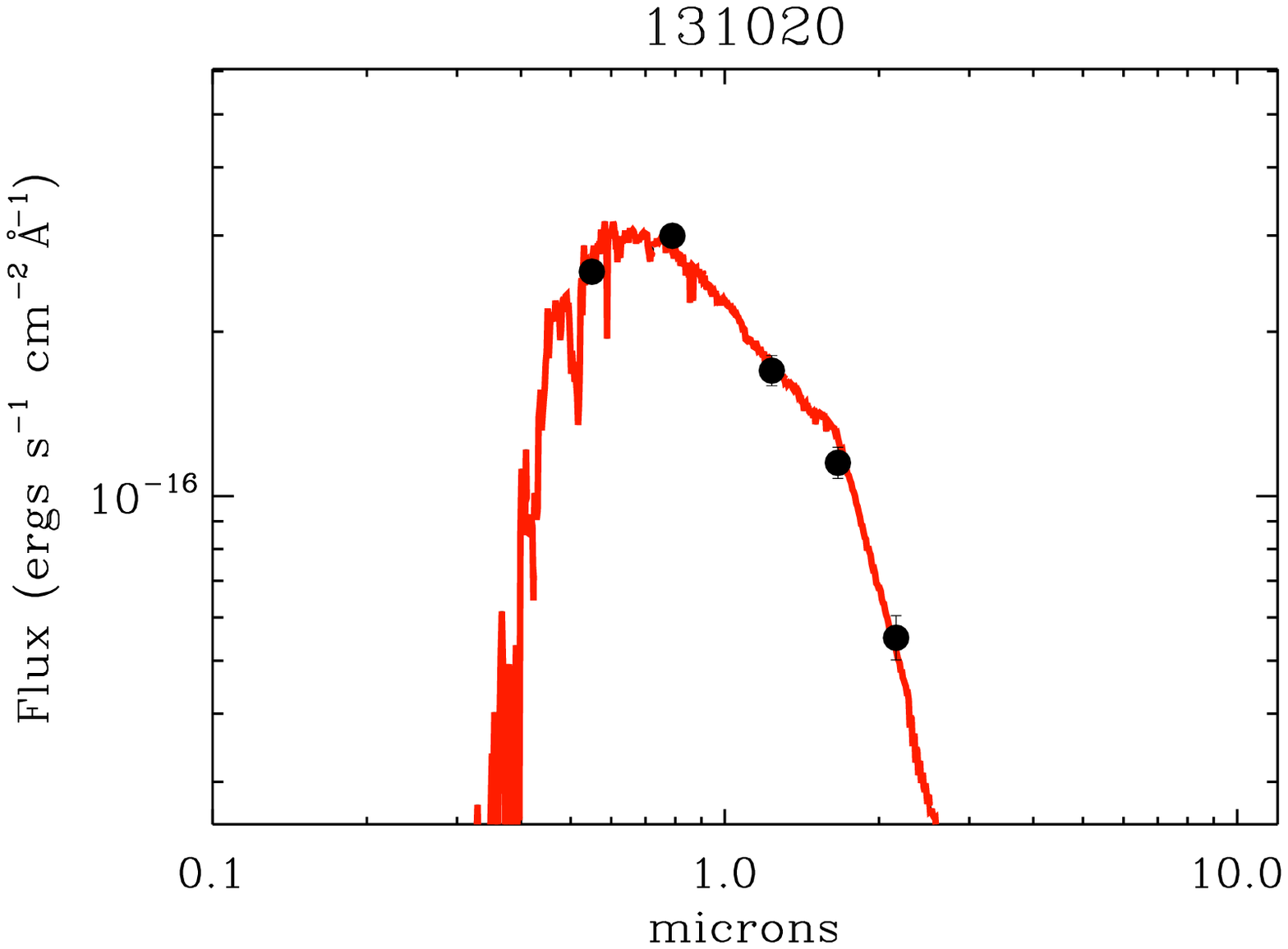}}
\subfigure{\includegraphics[width=.32\linewidth ]{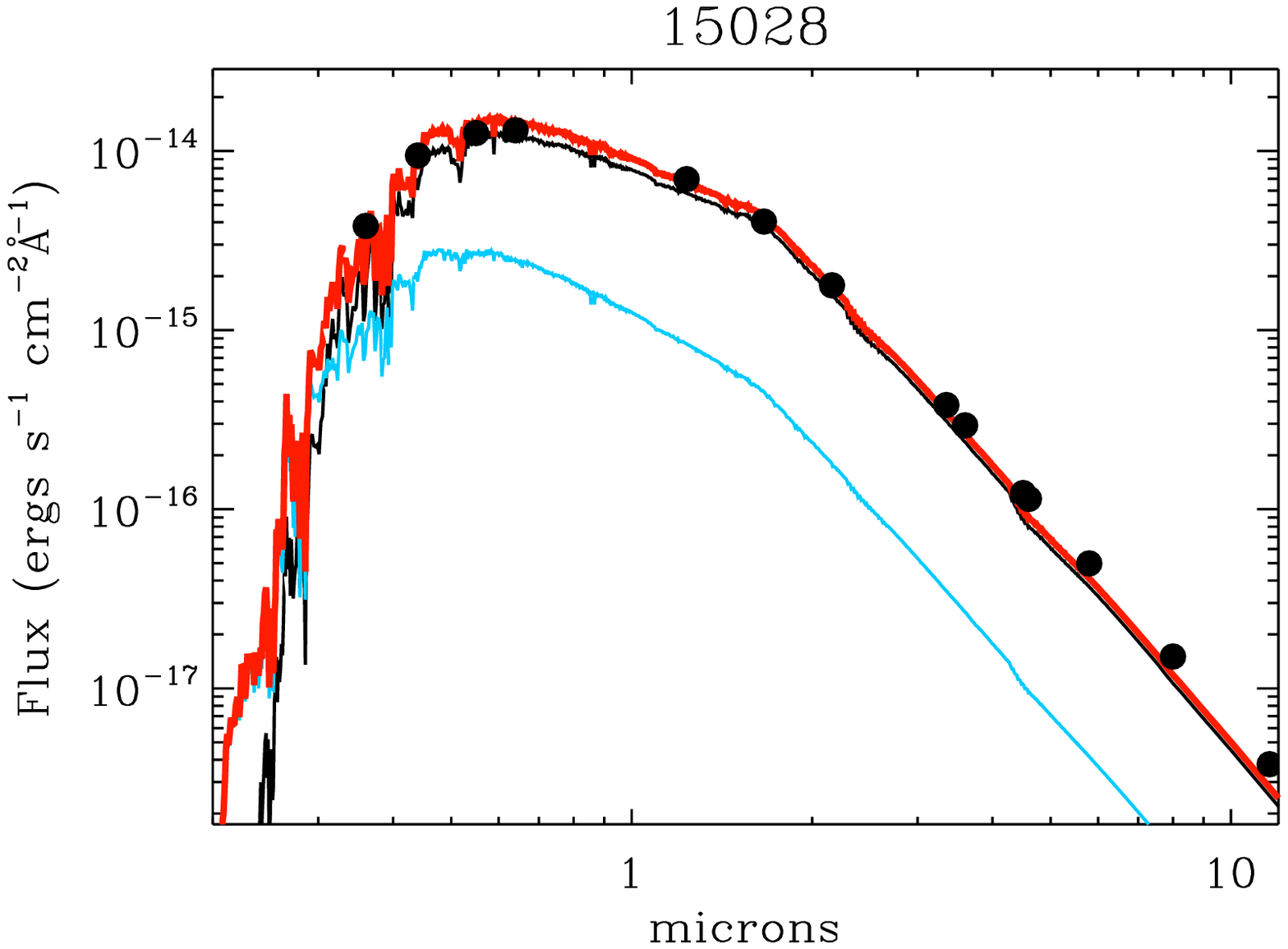}}
\subfigure{\includegraphics[width=.32\linewidth ]{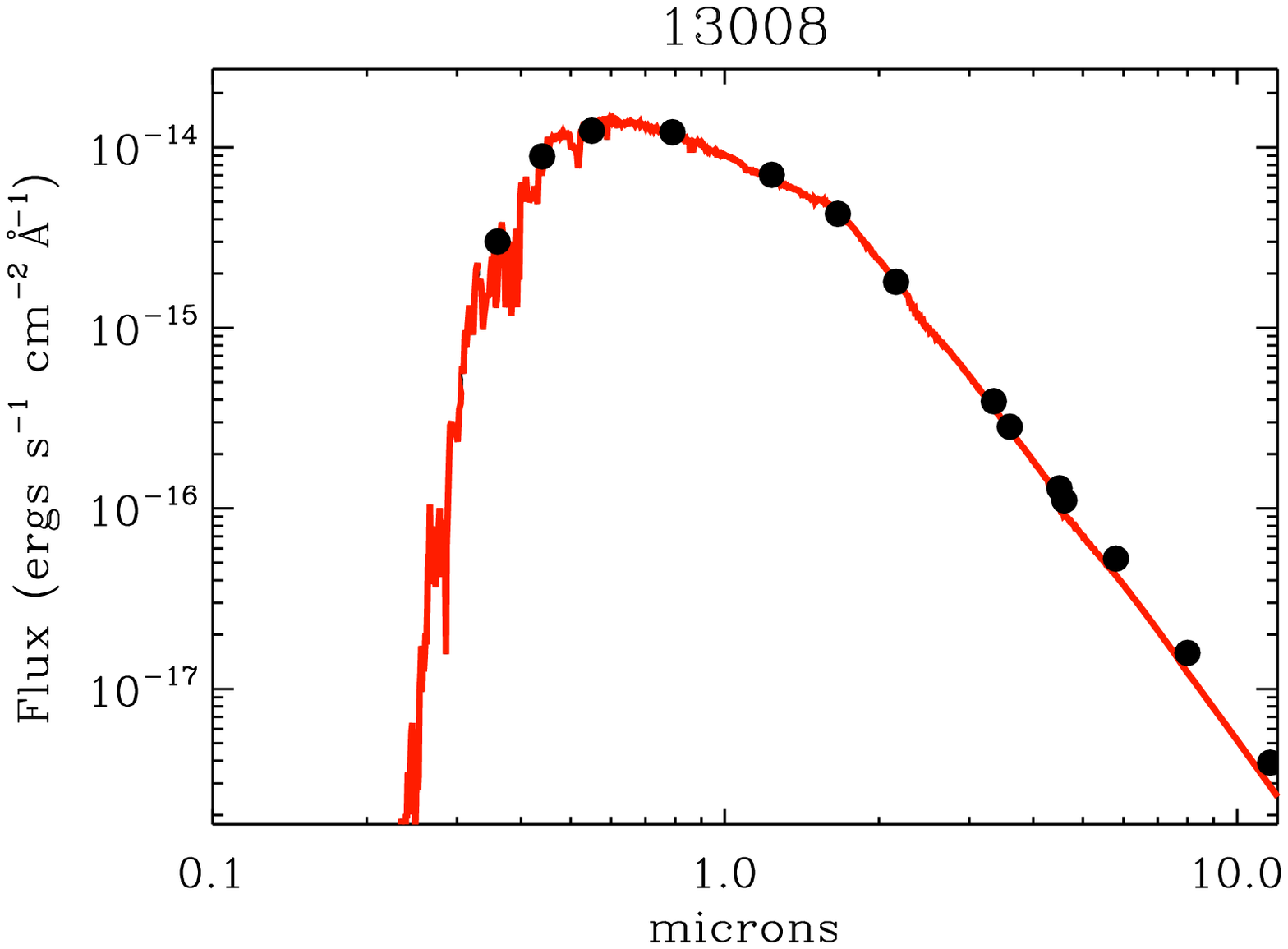}}
\caption{Best-fit SEDs for all six SSGs in M67 and NGC 6791. Observed flux is shown with filled circles. For the SB1s, we assume the flux contribution from the secondary to be negligible and show the flux from the primary in red. For the SB2, 15028, we assume a flux from a main-sequence secondary with R=0.83 and T$_\text{eff}=5250 $K based on the analysis of \citet{Mathieu2003}. For this star we show the flux contribution from the secondary in blue, the contribution of the primary in black, and the combined light in red. }\label{SEDs}
\end{center}
\end{figure*}

We performed a $\chi^2$-minimization between the observed photometry and a grid
of Castelli-Kurucz models \citep{CastelliKurucz} convolved with filter transmission functions. We fit only T$_\text{eff}$ (K) and R (\Rsolar) while
fixing the distance and reddening to cluster values. Altering the distance does significantly affect the values of radius and bolometric luminosity we determine, and so we ran our code using a range of distance values (3900-4100 pc for NGC 6791; 800-900 pc for M67) found in the literature to better determine the range in radius and luminosity. 

Photometry used in these SEDs is listed in Table \ref{tab:photometry}. Because 5 of the 6 systems are known photometric
variables, the larger source of error in some bandpasses is the intrinsic variability of the system and not the photon
statistics. The amplitude of this variability ranges from a few percent in the V band, up to 0.26 mags for WOCS 130013, the most variable SSG \citep{deMarchi2007, Mochejska2005, vandenBerg2002}. In order to fit the SED we therefore use the amplitude of the variability of the star rather than the photometric errors for the optical photometry. This variability is known for each star in at least the V band. Where the amplitude of variability is known, we use that as our error. If we do not have a measurement of variability in an optical band, we use the V-band variability. Spot modeling predicts that IR observations are much less effected by spot
modulation, and therefore for 2MASS, WISE, and Spitzer observations we use the photometric errors for the SED fits. In most cases we expect the amplitude of the variability in these bands to be less than the photometric errors.

Fit parameters for each of the SSGs are given in Table 3. For comparison, we also used our code to fit subgiants near the base of the RGB in NGC 6791 and M67: WOCS 10006 in M67 and WOCS 12270 in NGC 6791. 

These fits suggest that SSGs are slightly cooler and larger than a typical cluster subgiant. The SSG radii would place them on the lower RGB or near the end of the subgiant branch, but with cooler temperatures and lower luminosities than expected for a typical cluster giant. 

We note that a few of the stars show tentative evidence of an IR excess. However, given the large uncertainties on some of the WISE photometry and the variable nature of the stars, its not clear that this excess is significant.


\begin{deluxetable*}{l l c c c l}
\tabletypesize{\footnotesize}
\tablewidth{\linewidth}
\centering
\tablecaption{SED Best-fit Parameters for SSGs and Subgiant Comparison Stars}\label{fitparams}
\tablehead{\colhead{Cluster} & \colhead{WOCS ID} & T$_\text{eff}$ (K) & R (\Rsolar) & L
  (\Lsolar) & Type}
\startdata
M67 & 15028 & 4500 & 2.5-2.9 & 2.32-3.13& SSG\\
M67 & 13008 & 4500 & 2.8-3.1 & 2.92-3.57&SSG\\
NGC 6791 & 130013& 4250 & 2.9-3.2& 2.50-3.03&SSG\\
NGC 6791 & 147014 & 4500 & 2.3-2.5 & 1.97-2.32&SSG\\ 
NGC 6791 & 170008 & 4750 & 1.9-2.1 & 1.67-2.03&SSG \\
NGC 6791 & 131020 & 4250 & 2.3-2.5 & 1.56-1.85 &SSG\\
& & & & \\
\hline
\hline
\\
M67 & 10006 & 5250 & 2.4-2.7 & 3.97-5.02 & Subgiant\\
NGC 6791 & 12270 & 5000 & 1.9-2.1 & 2.04-2.50& Subgiant\\
\enddata
\end{deluxetable*}


\section{A Mass-transfer Origin for Sub-subgiants}

One hypothesis for SSG formation is that Roche lobe overflow reduces the mass of a subgiant star, lowering its luminosity and moving it into the SSG CMD region. In order to
investigate this idea we employ two different stellar evolution codes:
Binary Star Evolution (BSE; \citealt{Hurley2002}) and Modules for Experiments in Stellar
Astrophysics (MESA; \citealt{Paxton2011, Paxton2013, Paxton2015}). We use BSE as
an efficient tool to search the large progenitor-binary parameter space. We use MESA to produce more detailed models
of the evolution of systems BSE indicates may produce SSGs. 

\subsection{BSE Mass Transfer Models\label{section:BSE}}

\subsubsection{Genetic Algorithm}
We first used BSE to simulate binaries in clusters with parameters matching those
of NGC 6791 and M67 (see Section 2.1).The genetic algorithm creates 100 generations of 5000 binaries each, and for each cluster we perform 20 simulations. To begin, we define a sample of 5000 binaries for the first generation with: 

\begin{itemize}
\item primary masses chosen randomly from a uniform distribution between 0.7 \Msolar~ (well below the MSTO in both clusters) and twice the turnoff mass of the cluster,
\item secondary masses chosen randomly from a uniform distribution between 0.1 \Msolar~ and twice the turnoff mass,
\item periods chosen randomly from a uniform distribution between 3 days and 5000 days, 
\item and eccentricities chosen randomly from a uniform distribution between 0 and 1.
\end{itemize}
These distributions cover the relevant initial parameter space, but are not meant to reproduce the true shapes of these binary distributions, for example as observed in open clusters.

BSE then evolves these 5000 systems up to the age
of the cluster (4 Gyr for M67, 8 Gyr for NGC 6791). We use the default parameters from BSE, but we make two changes to the code:
 1) We increase the strength of the convective tidal damping coefficient by a factor of 100 to correspond with the findings of \citet{Geller2013}, and 
 2) we fix a bug in the implementation of Equation 32 in \citet{Hurley2002}\footnote{Specifically, we change  f=MIN(1.d0, (ttid/(2.d0*tc)**2)) to
   f=MIN(1.d0, (ttid/(2.d0*tc))**2) in two locations in evolv2.f}

Once the systems have been evolved to the age of the cluster, we evaluate the fitness of each model. We evaluate fitness based on two criteria: the observed location of a system in a BV CMD, and the period of the final binary system. Specifically, we define the fitness (F) as: 
\begin{equation}
F=f_{reg}f_{BV}f_{V}f_{p}
\end{equation}
where
\begin{equation}
f_{reg}=\begin{cases}1 &\mbox{\tiny{Star falls redward of the equal-mass binary sequence}}\\ 0 &\mbox{\tiny{Otherwise}}\\
\end{cases}
\end{equation}
and 
\begin{equation}
f_i=e^{\frac{-(O_i- S_i)^2}{\sigma_i^2}}
\end{equation}where i=B-V, V, or P, respectively, O refers to the observed color, magnitude or period of the SSGs, and S the color, magnitude, or period of the BSE model. 

For M67, we took $O_{B-V}=0.9$ and $O_V=13.8$. For NGC 6791 we took $O_{B-V}=1.25$ and $O_V=17.7$. For both clusters, we select for short period systems by taking $O_{P}=10$ days, and we adopt $\sigma_{P}=3$ days, $\sigma_{B-V}=0.15$, and $\sigma_{V}=0.3$. 

For a second round of models, we sought to produce systems matching the orbital
periods of the SSGs in NGC 6791 and M67. For this we re-ran the
genetic algorithm for each cluster, this time taking $O_{P}$ to be the specific orbital period of each SSG. For 15028, the SB2 SSG, we also included a fitness term for the mass ratio of the system ($\frac{M_2}{M_1}=q$) with $O_{q}$=0.7. 

After evaluating the fitness of each of the 5000 first-generation models, we take any models with non-zero fitness and these models become ``parents" for the next generation of models. Technically we limit the number of parents per generation to 1000, but we rarely find more than a few hundred.  Parents are then allowed to ``mate" with each other to produce two ``children'' per parent-parent pair in the next generation.  To define the children, we begin with the parent that has the highest fitness value, allow it to mate with all other parents, repeat this process for the parent with the second highest fitness value, and so on until we obtain all parent-parent combinations or we produce 3000 children.   

To produce a child we take a random combination of the initial parameters from each parent.  Specifically, to determine each initial binary parameter for a child, we draw a random number from a uniform distribution between 0 and 1; if the number is $< 0.5$, we choose the initial parameter of the first parent, and otherwise we choose the initial parameter from the second parent for the child.  If a child duplicates a binary already in the subsequent generation, we impose a mutation where at least one of the binary initial parameters is chosen randomly from the same respective distribution defining the initial parameters if the first generation, and the number of mutated parameters is chosen randomly.

We take the $\leq 3000$ children produced in this manner, and fill the
remaining $\geq 2000$ spots with binaries whose parameters are chosen from the same distributions as the initial generation. We then evolve this new generation of 5000 binaries with BSE up to the cluster age, and the process is repeated for 100 generations.  Through this procedure, we ensure that subsequent generations climb to higher and higher fitness values, retain the best fitting binaries throughout the generations, and introduce a fresh sample of random binaries in each generation to fill out the parameter space.  

Note that the genetic algorithm does not uniformly sample parameter space, and therefore can potentially miss a peak in the fitness surface.  This is alleviated somewhat by the introduction of binaries with randomly chosen initial parameters into each generation.  We ran 20 simulations of the genetic algorithm (with different initial random seeds) and combine the results to further alleviate this issue. 	

\subsubsection{Results of BSE Genetic Algorithm}
In both sets of runs, those selecting for an orbital period of 10 days and those selecting for specific SSG orbital periods, BSE was able to produce systems in the SSG region of a CMD. Inspection of these results show that there is a family of solutions that
create these SSG systems. In M67, the progenitor binaries are generally high-eccentricity systems with periods between 3 and 1000 days, with the longest
period binaries requiring eccentricities approaching 1.  The secondary star can possess a
range of masses between 0.3 and 0.8 \Msolar, with the majority being drawn from the lower end of this range.  The primary star is a $\sim1.3$
\Msolar~ star that begins Roche lobe overflow somewhere on the subgiant
branch. This requires that tidal forces circularize and shrink the initial
orbit of these systems so that when they evolve through the subgiant branch
they have periods of $\sim1$ day. We note that our choice of a large tidal strength factor may
artificially allow the wide, high eccentricity systems to circularize, but
this does not change the conclusion that we require $\sim 1$ day 
binaries on the subgiant branch to create SSGs via mass transfer. 

The final SSG systems are short period (P $<6.75$ days) circular binaries. The highest fitness solutions have reduced the primary mass to $\sim$0.2 \Msolar, and increased the secondary mass to $\sim$1.0 \Msolar, though a wide range of final primary and secondary masses are produced by the algorithm. 

In NGC 6791, results are similar.  The progenitor binaries are again high-eccentricity systems with periods from 3 days up to 1000 days, with
longer period systems requiring higher eccentricities. Initial primary masses
are $\sim 1.1$ \Msolar, with secondaries ranging from 0.3-0.5 \Msolar, with
the majority drawn from the lower mass end. The final systems are circular
binaries periods of just
a few days. The longest period system created was a 3.26 day binary.

BSE was unable to produce SSGs
with periods above 3.5 days in NGC 6791, or 7 days in M67. These periods are shorter than 5 out of the 6 SSGs in our sample.The only individual
SSG that we had some success at reproducing using the genetic algorithm was
15028. This 2.8 day binary falls solidly in the period domain that can be
created by mass transfer. However, creating an SSG with the observed mass ratio of this system ($q=0.7$) proved problematic. The genetic algorithm strongly favored smaller $q$ values for longer period SSGs, with a $q=0.7$ possible only for shorter period binaries with $P \sim 1$ day. 

We note that the algorithm produces longer period SSGs in M67 than in NGC 6791,
which is as expected. The Roche lobe radius is given by \citet{Eggleton1983} as: 
\begin{equation}
\frac{r_{L}}{a}= \frac{0.49q^\frac{2}{3}}{0.6q^\frac{2}{3}+\text{ln}(1+q^\frac{1}{3})}
\end{equation} where $r_L$ is the Roche lobe radius, a is the orbital separation, and $q$ is the mass ratio of the binary system. 
The Roche lobe radius depends only on the mass ratio and the orbital separation of a binary. The stellar radius depends on a star's mass and evolutionary state.  Since subgiants in NGC 6791 are lower mass than in M67, they begin with smaller radii. Therefore, for a given mass ratio and orbital separation, NGC 6791 subgiants must reach a more evolved state to exceed their Roche lobes. This means they will have evolved further up the giant branch than their M67 counterparts, and thus not evolve through the SSG domain as they lose mass. In order for an NGC 6791 subgiants to evolve through the SSG region as they lose mass, they must then have smaller orbital separations than are required for M67 subgiants. 

The results indicate that, while we do expect mass transfer to create SSG
systems, we would expect the observed periods to be in the range of just a few
days in both clusters, shorter than most of the observed SSGs. BSE therefore
suggests that none of the NGC 6791 SSGs have short enough periods to be mass transfer
systems. Similarly, while mass transfer may create a 2.8 day binary like the one observed in M67, the expected mass ratio would be much smaller than the observed $q=0.7$, and BSE is unable to create the other 18.4 day binary.

\subsection{Detailed MESA Modeling}
The BSE-based genetic algorithm is an excellent tool for exploring a wide
range of parameter space, but the stellar evolution and mass transfer
calculations are highly parameterized. We therefore also use the  more
detailed evolution code Modules for Experiments in Stellar Astrophysics (MESA; \citealt{Paxton2013}) to create SSGs via stable mass transfer, and
compare the results of these MESA models to our observations.  

\begin{figure*}[htbp]
\begin{center}
\includegraphics[width=.8\linewidth ]{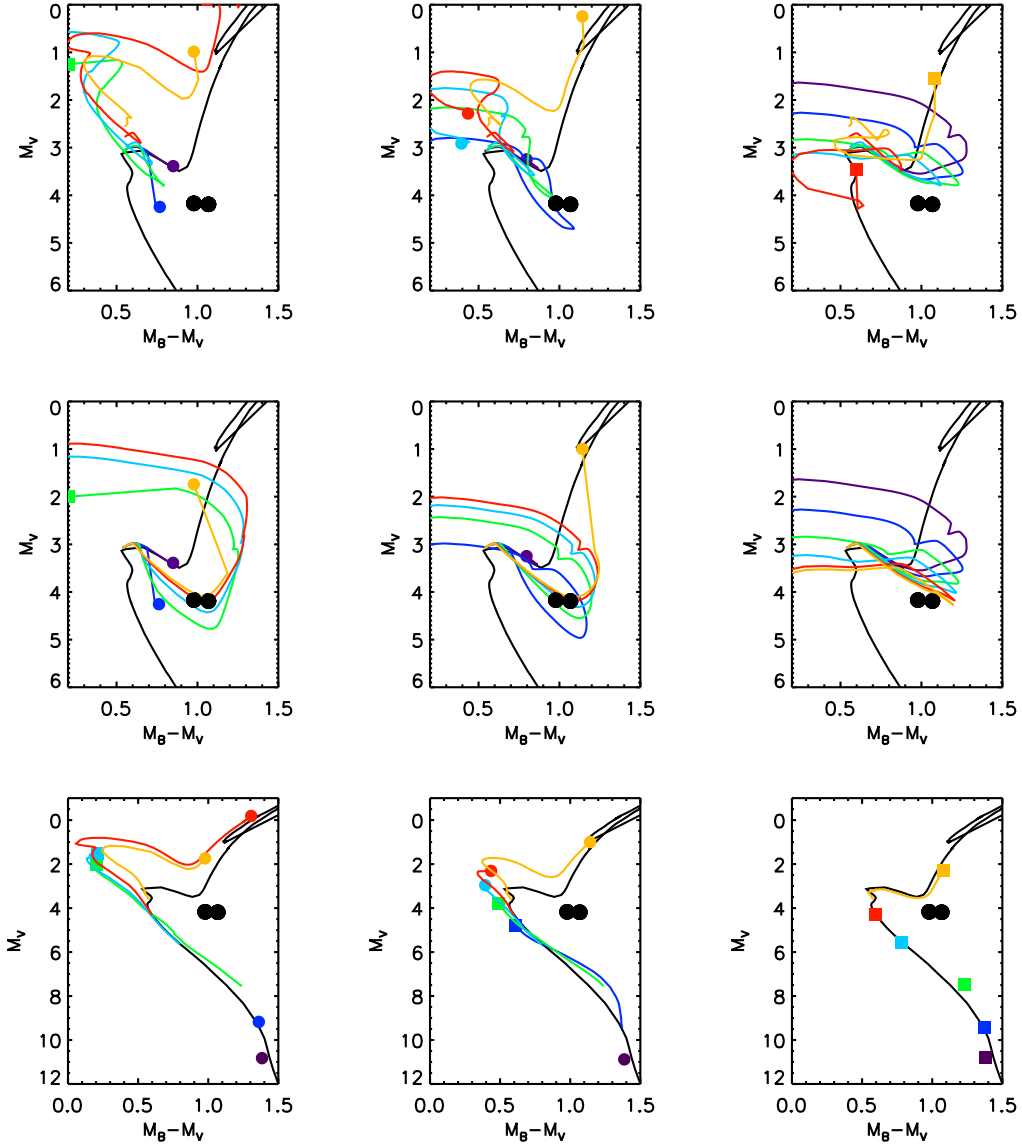}
\caption{Evolutionary tracks for mass transfer SSGs in M67 given three
  different mass transfer efficiencies. All models show a 1.3 \Msolar~subgiant
primary in a 1.0 day binary. Mass of the secondary for each system is
indicated by color of the track: 1.25 \Msolar~(yellow), 1.1 \Msolar~(red), 0.9
\Msolar~(light blue), 0.7 \Msolar~(green), 0.5 \Msolar~(dark blue), 0.3
\Msolar~(purple). The first column displays conservative mass transfer, the
middle column shows 50$\%$ efficient mass transfer, and the last column shows
0$\%$ efficient mass transfer. The top row shows the evolution of the combined
light of the system, the middle row shows the evolution of the primary
only, and the bottom row shows the evolution of the secondary. Mass transfer
tracks were all evolved up to 5 Gyr, except in the case the model was
terminated due to the onset of dynamically unstable mass transfer and/or common envelope evolution. The colored symbols indicate the end of the evolutionary track, either at 5 Gyr (filled square) or due to the onset of unstable mass transfer (filled circles). The black circles indicate the location of the SSGs. A 4 Gyr isochrone is shown in black \citep{Bressan2012}.}\label{fig:MESAgrid}
\end{center}
\end{figure*}

For M67 we initially ran a coarse grid of models with input based on the BSE results . These models all had a
primary mass of $M_1=1.3$ \Msolar~and companion masses in the range $0.3$
\Msolar $< M_2 < 1.25$ \Msolar. We evolved each component of the binary up to
core hydrogen exhaustion before placing it in a binary with a period between 1
and 10 days and allowing the evolution to proceed. We start from the test suite case \texttt{binary\char`_both\char`_stars}, use non-rotating models, and do not include magnetic
braking. We used three different mass transfer
efficiencies: $\alpha=0.0, 0.5$, and $1.0$. $\alpha$ is defined such that
fully conservative mass transfer has $\alpha=0.0$, and if no mass is
transfered $\alpha=1.0$. Mass and angular momentum
are presumed to be lost from the vicinity of the primary in these models. 

Models with periods above $\sim 2$ days began Roche lobe overflow after
beginning their ascent up the red giant branch, and mass transfer products did
not evolve through the SSG region. In many cases mass transfer was dynamically unstable
and we terminated their evolution, as MESA cannot handle these cases. It is generally assumed that binaries undergoing unstable Case B mass transfer
enter a common envelope phase that ends with the spiraling in of the binary
and an ejection of the common envelope material \citep{Paczynski1976, Ivanova2013}. The end product of this
phase is either a shorter period white dwarf-main sequence binary or a merger
to create a single star. If this is correct, we conclude these longer period
systems do not create SSGs.

Models with periods
less than $\sim 2$ days began Roche lobe overflow while still on the subgiant
branch and did proceed through the SSG region as mass transfer proceeded. For these shorter period models
we ran a more detailed grid of models with periods between 0.6 and 2.0 days,
$M_1=1.3$  \Msolar, varying $M_2$ from  $0.3$ \Msolar $< M_2 < 1.25$ \Msolar, and using three mass
transfer efficiencies  $\alpha=0.0$, $0.5$, and $1.0$. Our models
indicate that systems with initial periods $P< 0.8$ days begin RLO prior to
evolution onto the subgiant branch and do not pass through the SSG
region. Models with $0.8$ days$ < P \leq1.2$ days will often evolve through the
SSG region depending on $M_2$ and mass transfer
efficiency. Models with  $P > 1.2$ days may have primaries that evolve through the SSG region, but mass transfer is only stable in these systems if they have a near equal-mass secondary. Due to the required mass of the secondary, the combined light of the binary does not pass into the SSG domain. Therefore, it appears that
only a very narrow range of systems with periods right around $P=1.0$ day
begin mass transfer at the right time in their evolution, with faint enough companions to move through the SSG domain. 

As an example , we show a grid of MESA models with a 1.0 day period, 1.3 \Msolar~primary and a range of
secondary masses from 0.3 \Msolar~to 1.25 \Msolar~in Figure~\ref{fig:MESAgrid} with
the secondary mass indicated by the color of the track. In this grid, for a large range of secondary masses and mass transfer
efficiencies, the primary will evolve near the domain of the M67 SSGs (see Figure 3, middle row. However,
only for models beginning with a fairly low-mass secondary and a moderate
degree of non-conservativeness will the combined light of the binary evolve
through this region (Figure 3, top row). For example, see the top middle plot. Only the model binaries with
progenitor secondaries of 0.5 and 0.7 \Msolar~are observed in the SSG
region during their evolution.

The evolution of the accreting stars in these models is also interesting. We assume the accreting star in these binaries is a main sequence star with initial mass ranging from 0.3 to 1.25 \Msolar. For models where mass transfer can proceed stably and we assume a significant fraction of the mass lost from the primary is accreted by the secondary, the secondaries may gain several tenths of a solar mass of material. This accretion causes them to move up the main sequence to a position corresponding to their new larger mass. If the accreting star starts close enough to the main sequence turn off, or we assume a large mass-transfer efficiency, the accreting star may move above the main-sequence turn-off and in to the blue straggler region (for example, see Figure 3, lower left plot). As these accretors begin to evolve off the main sequence, their higher mass causes them to follow an evolutionary track brighter and bluer than the cluster isochrone. Such `yellow stragglers' have been observed in M67 and other clusters, and are believed to be evolved blue stragglers \citep{Landsman1997, Leiner2016, Geller2015}. In this mass transfer scenario, then, SSGs could be part of the same evolutionary pathway that leads to the formation of other non-standard stars like the blue and yellow stragglers.

We ran another grid for NGC 6791, using a primary of $M_1=1.1$ \Msolar, varying  $M_2$ from  $0.3$ \Msolar~$< M_2 < 1.0$ \Msolar, using three mass transfer efficiencies  $\alpha=0.0$, $0.5$, and $1.0$,  and periods between 0.6 and 10.0 days. Results were similar to those of M67. The models indicate that systems with initial periods $0.6 \leq P \leq 1.0$ days moved through the SSG region during RLO. As in M67, longer period systems began mass transfer on the lower giant branch, and if mass loss proceeded stably the model evolved to the red of the RGB, not down into the SSG region. 

\begin{figure*}[htbp]
\subfigure{\includegraphics[width=0.5\linewidth ]{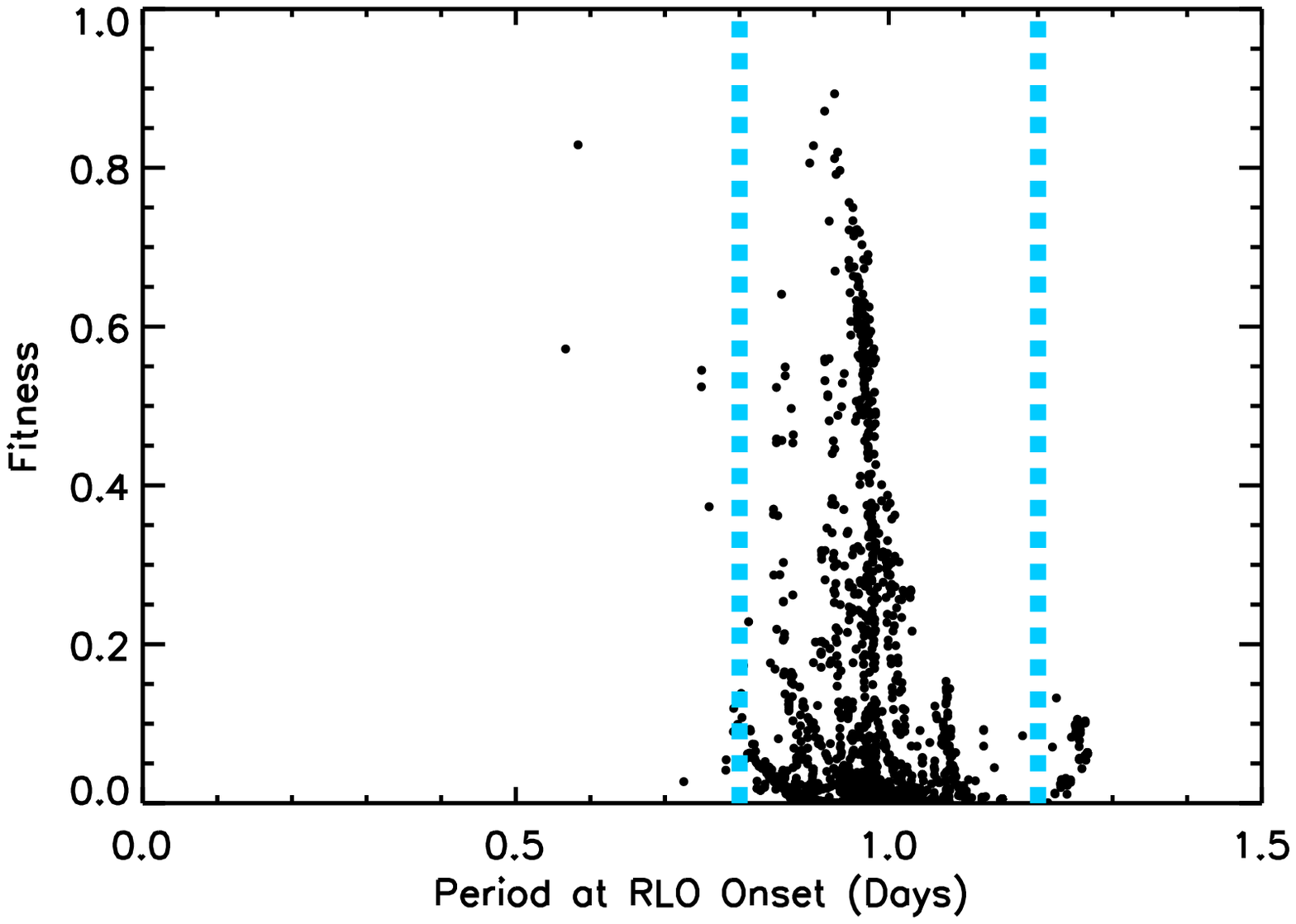}}
\subfigure{\includegraphics[width=0.5\linewidth ]{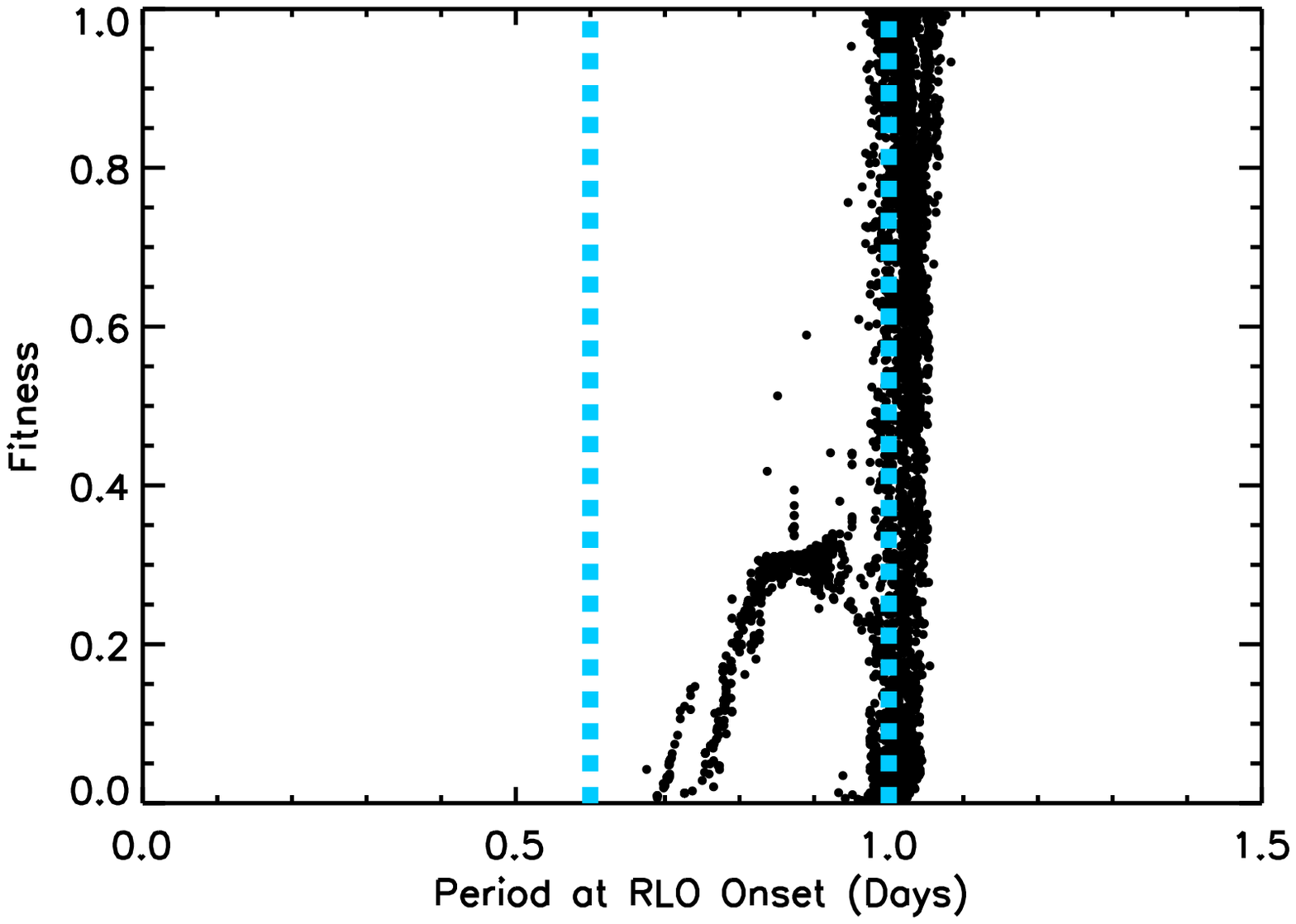}}
\subfigure{\includegraphics[width=0.5\linewidth ]{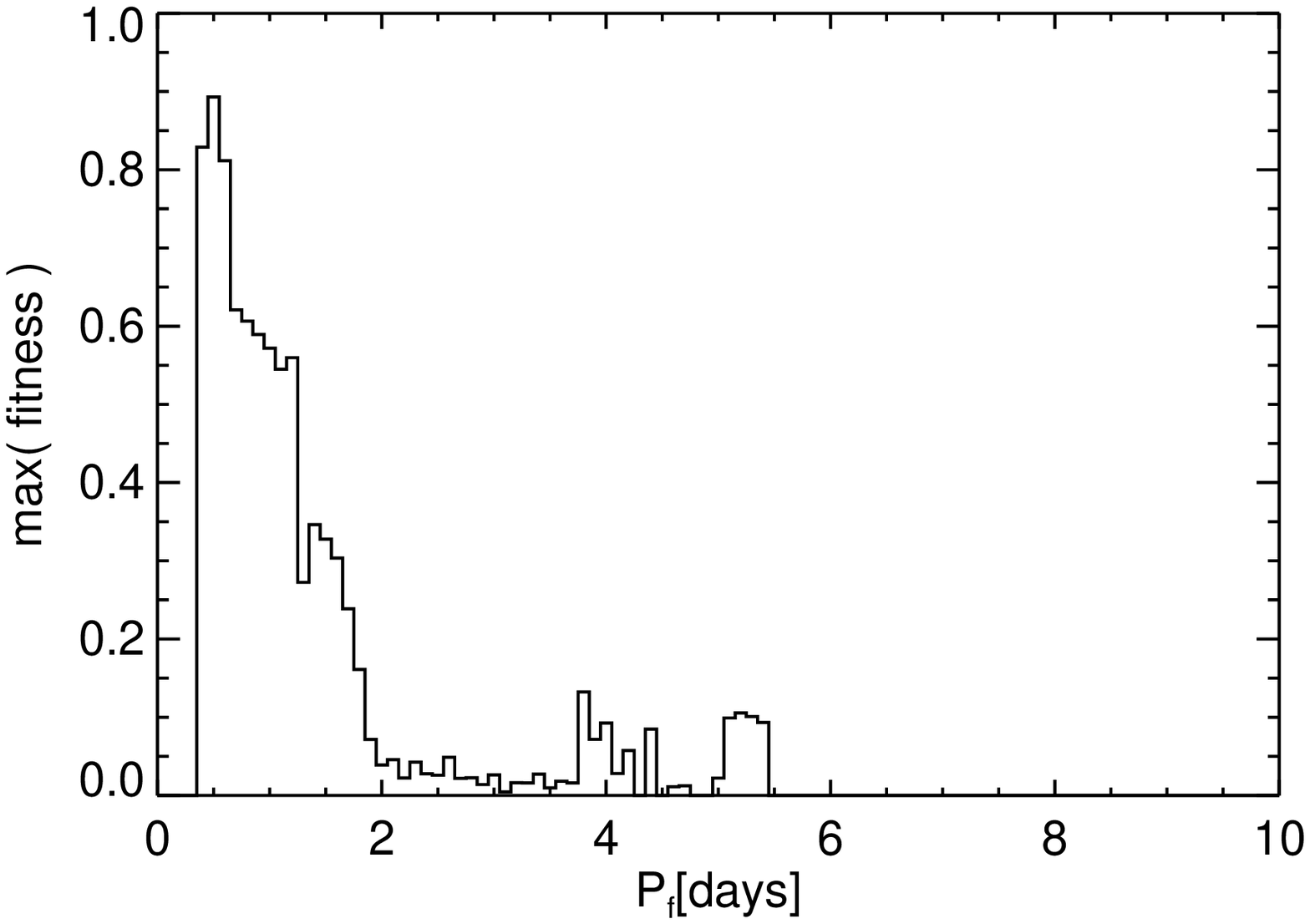}}
\subfigure{\includegraphics[width=0.5\linewidth ]{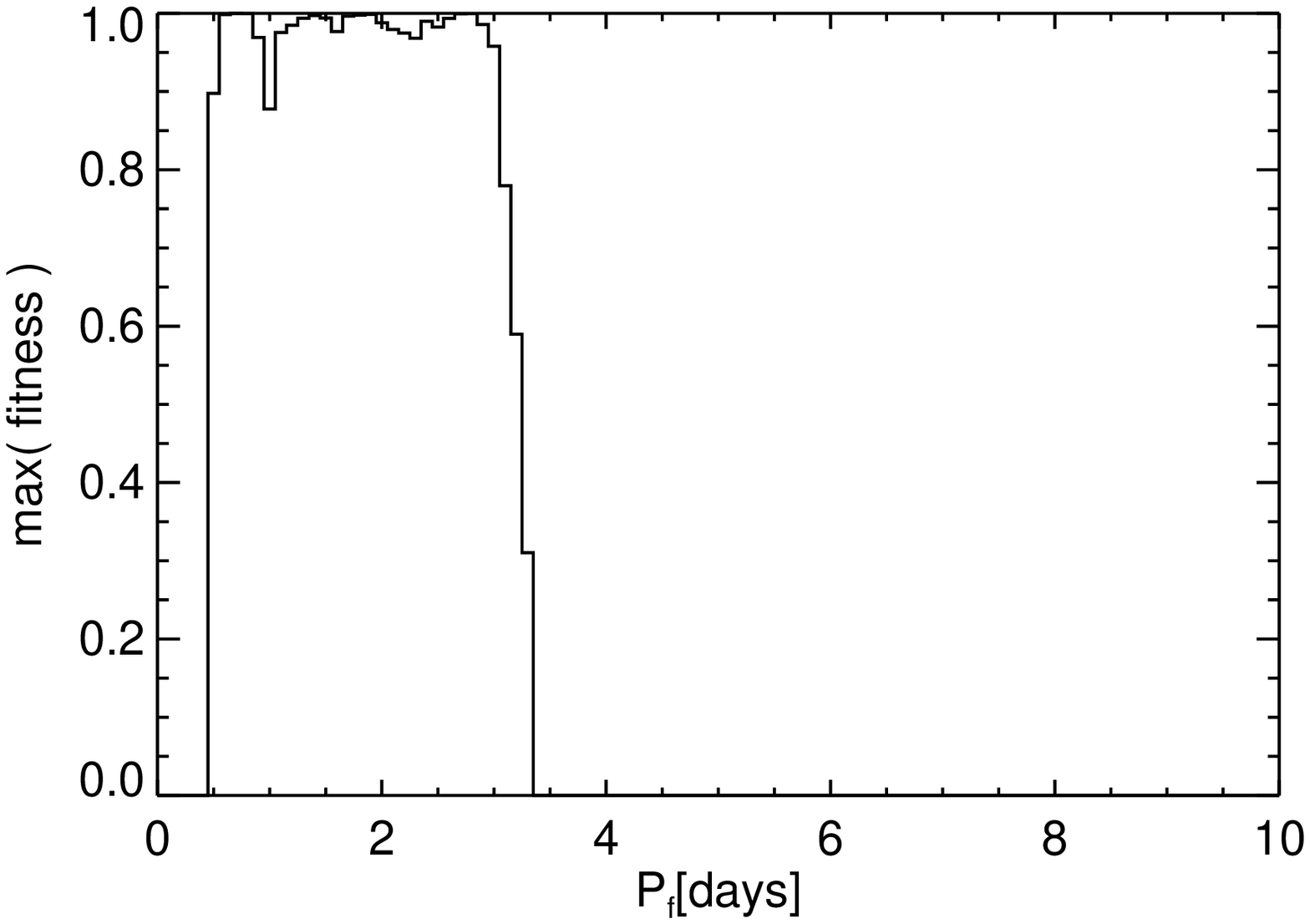}}
\caption{\textbf{(top)}Plots of the fitness of BSE SSG models versus the orbital period at onset of RLO in M67 (top left) and NGC 6791 (top right).  The longest and shortest periods that produce SSGs in MESA are also overplotted with dashed blue vertical lines. In both clusters, the highest fitness SSGs are all produced by systems with $\sim 1$ day orbital periods.\textbf{(bottom)} Plots of the maximum fitness of BSE SSG models versus the model SSG's final orbital period at 4 Gyr for M67 (bottom left) or 8 Gyr for NGC 6791 (bottom right) }\label{Period_histograms}
\end{figure*}

We compare this finding to the results of the BSE models. Shown in
Figure~\ref{Period_histograms} is a histogram of orbital periods of the SSG binaries
produced in BSE models of M67 and NGC 6791 (Section
\ref{section:BSE}). We show both the orbital period of the system at onset of RLO, and the final period of the binary at 4 Gyr when it is observed as an SSG. We also show the shortest and longest period SSG created with MESA with vertical dashed blue lines. Note that our grid of MESA models has a resolution of 0.1 days in period. We conclude that the MESA results are consistent with our findings from BSE that only binaries in $\sim 1$ day orbits at onset of RLO will move through the SSG region. 

\subsection{Frequency of Mass Transfer Formation}\label{section:frequency}
Given this period range for SSG formation, we can use the period distribution
found by \citet{Raghavan2010} to estimate the number of predicted SSG systems
in a cluster. Specifically, \citet{Raghavan2010} fit a Gaussian function to
the distribution of orbital periods found in a large sample of binary
systems. They find $\mu_{log P}$=5.03 and $\sigma_{logP}=
2.28$. Integrating this function using the upper and lower period bounds found for M67 ($0.8 \leq P \leq 1.2$) and NGC 6791 ($0.6 \leq P \leq 1.0$) we would expect only $\sim0.3\%$ of binaries to go through this evolution. Assuming a $50\%$ binary fraction and given a population of  $\sim30$ subgiant stars in M67, and $\sim100$ subgiant stars in NGC 6791 (see Section 6.4), this results in a Poisson probability of observing one or more mass transfer SSGs in M67 of 4$\%$ and  in NGC 6791 of $14\%$. This estimate should be regarded as an upper limit, as period is not the
only factor which determines whether a star moves through the SSG region. For
example, if the companion mass is very small, mass transfer may be unstable
and the system would not evolve as an SSG. Conversely, if the companion is too
close to the main sequence turn off, the secondary may overwhelm the
lower-luminosity primary and move the system into a more standard region of
the CMD or into the blue straggler domain. 

Given this small number, it is unlikely, though not impossible that we would observe
a mass-transfer SSG in NGC 6791 or M67. However, this mechanism is unlikely to explain all
the SSGs observed in M67 or NGC 6791. The Poisson probability of producing 2
SSGs in M67 or 4 in NGC 6791 from mass transfer is negligible. 

However, in a larger cluster it may be quite likely to observe at least one mass transfer SSG. In a companion paper, Geller et al. 2017b (in preparation), we investigate in more detail the expected formation frequency of this and other formation mechanisms across a wide range of cluster properties. 

\subsection{Tidally Enhanced Wind}
A serious mismatch between mass transfer models and observations is that model
mass transfer SSGs have shorter periods than most SSGs observed in M67 and NGC 6791. One method to produce longer period binaries undergoing mass loss is to adopt
a model in which the primary can lose substantial mass via a wind while still well within
its Roche lobe. 
The tidally enhanced wind model proposed by \citet{Tout1988} proposes that tidal interactions and magnetic activity drive a stronger stellar wind in close binary systems than in a typical single star. They assume that the wind can be described by the standard Reimer's wind for RGB stars, multiplied by a factor that has the same dependence on stellar radius ($R$) and Roche lobe radius ($R_L$) as a tidal torque. Specifically, their expression is:

\begin{equation}
\begin{aligned}
\dot{M}_{Wind}=\dot{M}_{Reimers} \times (1+B\times min[\frac{R}{R_L}^6, \frac{1}{2^6}])\\
\mbox{where } \dot{M}_{Reimers}=-4 \times 10^{-13} \bigg( \frac{RL}{M}\bigg)
\end{aligned}
\end{equation}where R, L, and M are in solar units and time is in years. Here they assume that wind mass loss saturates when $\frac{R}{R_L}=\frac{1}{2}$. This wind prescription includes a constant multiplicative factor ($B$) that may be varied to achieve greater or lesser mass loss rates. \citet{Tout1988} calibrate this constant to match the properties of the system Z Her, a detached RS CVn binary with a mass ratio inversion in which the more evolved star is near the end of the subgiant branch and is less massive than its near-turnoff companion. Its orbital period is $P=4$ days. They find $B=10^4$ well matches the observed mass loss from the primary. 

Using a tidally
enhanced wind model can reproduce an SSG similar to 15028 (Figure 5). This model has an initial primary mass of 1.3 \Msolar, a period of 2.8 days, and a coefficient of $B= 2 \times 10^4$, twice as large as that proposed in \citet{Tout1988}. The mass loss rates on the subgiant branch
required are on the order of $10^{-9}$ \Msolar~yr$^{-1}$.  The star has a mass of just 0.95 \Msolar~when it reaches the CMD location of
15028. 

While we have no direct mass measurement of 15028, we do have a mass ratio of $q=0.7$ from the orbital solution. Given this mass ratio, the 0.95 \Msolar~mass from the tidally-enhanced wind model would imply a secondary of $0.67$ \Msolar, in which case the observed secondary is substantially hotter and more luminous than expected for a $0.67$ \Msolar~star \citep{Mathieu2003}. Alternatively, assuming the rotational and orbital axes are aligned, a secondary mass of $\sim0.9$ \Msolar~well matches the photometry, spectroscopic temperature, and mean density of the secondary star \citep{Mathieu2003}. This would imply a mass of 1.3 \Msolar~for the primary, indicating a subgiant that has not lost substantial mass. However, the luminosity ratio of the system is not consistent with the alignment of the axes, and \citet{Mathieu2003} were not able to find a fully self-consistent model for the system.

The SSGs in NGC 6791 are in longer-period orbits than 15028. They also presumably start with smaller radii if they are normal cluster subgiants undergoing mass loss. Therefore, the tidal wind enhancement does not produce a noticeable effect in the models until the stars have evolved substantially up the giant branch. Even increasing the B parameter by a factor of 10 is unable to
create observed systems near the location of the NGC 6791 SSGs. Similarly,  a tidally enhanced wind model for the 18.4 day binary in M67 also does not produce significant mass loss until the primary is substantially more evolved. These stellar models never move through the SSG region. 

Overall the wind prescription of \citet{Tout1988} is unable to reproduce the CMD
location of any of the NGC 6791 SSGs or the 18.4 day SSG in M67 using a value of B close to what is typically assumed. These stars are just not close enough to their Roche radii to 
have large mass loss rates using this model. This wind prescription can create the 2.8 day SSG in M67 by
losing $\sim 0.4$ \Msolar~from a subgiant primary,
but it is not clear from the observational evidence that this star has lost substantial mass. We conclude that wind mass loss
rates are likely not large enough to be the sole reason for the SSGs'
under-luminosity.

\begin{figure}
\begin{center}
\scalebox{1.0}{\includegraphics[width=0.9\linewidth ]{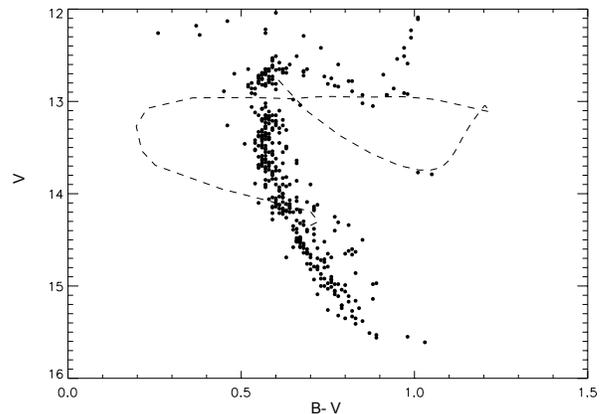}}
\caption{MESA evolutionary track showing the evolution of the combined light of a binary system with a tidally enhanced wind \citep{Tout1988}. The model has a 1.3~\Msolar~primary, a 0.9~\Msolar~ secondary, orbital period of 2.8 days, and a wind coefficient, B, of $2 \times 10^4$.  We assume no mass gets transferred to the secondary. This model evolves a star through the SSG region, passing through the area occupied by the two M67 SSGs.}\label{m67_enhancedwind}
\end{center}
\end{figure}

\section{Sub-subgiants from Envelope Stripping}
Another possibility is that SSGs could be created as a result of removing the envelope of a subgiant star. Rapid envelope mass loss yields a rapid decrease in luminosity, and subsequent evolution below the current subgiant branch to a lower-mass red giant branch. 

Such stripping could occur in a number of ways. One suggestion is that such mass loss may occur if a subgiant star has a close encounter with a passing star. This may occur, for example, during a resonant binary encounter (e.g. \citealt{Heggie1975, Bacon1996}). If the impact parameter of the passage is sufficient to disrupt  and remove a large fraction of the stellar envelope, but
not close enough to lead to a merger, an SSG-like star might result. 

\subsection{MESA models of subgiant mass loss}
In order to explore the effect of envelope mass loss on a subgiant, we remove mass from a subgiant star at a high, constant
rate using MESA, stopping the mass loss after the star has
lost a few tenths of a solar mass of material from its envelope and allowing
the star to continue to evolve without further mass loss. In principle, any mass loss rate in which the subgiant can lose a few tenths of a solar mass of material within its subgiant lifetime can move the subgiant into the SSG domain. This requires mass loss rates $\gtrsim 10^{-8}$ \Msolar~$ \text{yr}^{-1}$. In practice, MESA does not handle dynamical mass loss, and thus the highest mass loss rate for which we achieve numerical stability is $10^{-5}$ \Msolar~\text{yr}$^{-1}$.

\begin{figure*}[htbp]
\subfigure{\includegraphics[width=0.5\linewidth ]{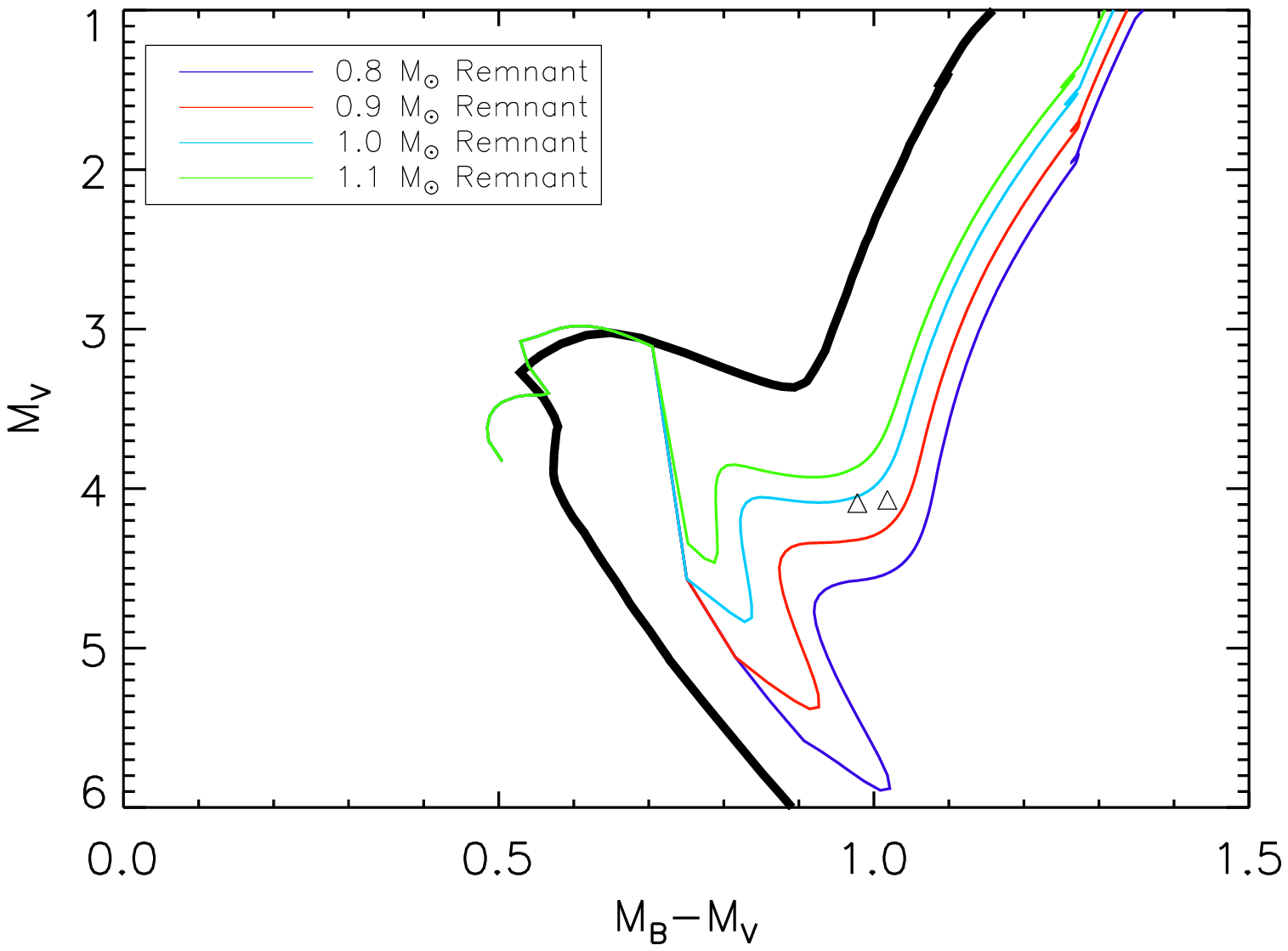}}
\subfigure{\includegraphics[width=0.5\linewidth ]{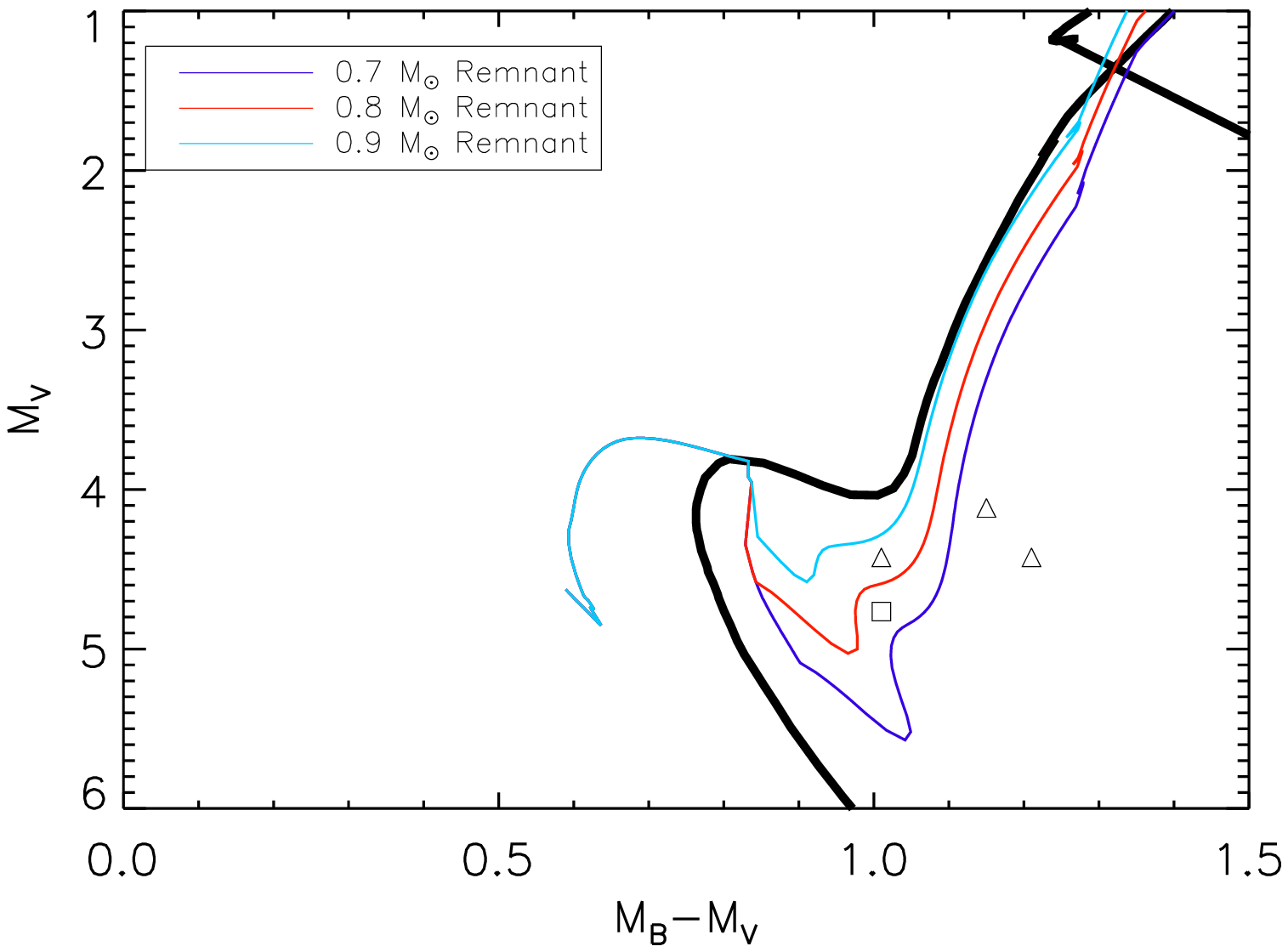}}
\caption{MESA evolutionary tracks showing mass rapidly removed from a subgiant star. \textbf{(left)} A 1.3 \Msolar~subgiant, mass loss rate of $10^{-5}$ \Msolar~yr$^{-1}$. Mass loss is terminated
  when the subgiant star has reached 0.8-1.1 \Msolar~ and the star is allowed to
  evolve normally. The plot shows the  evolution of this system in colored tracks, with a 4 Gyr Padova isochrone shown in black.  The triangles show the locations of the M67 SSGs. \textbf{(right)} A 1.1 \Msolar~ subgiant,
  mass loss rate of  $10^{-5}$
  \Msolar~yr$^{-1}$. Mass loss is terminated when the subgiant star has reached
 a mass between 0.7 \Msolar~ and 0.9 \Msolar~ and the star is allowed to evolve
  normally. The plot shows the evolution of this system in colored tracks, with an 8 Gyr Padova isochrone shown in
  black.  The triangles show the locations of the NGC 6791 binary SSGs. The
  square is the NGC 6791 single SSG. }\label{NGC6791_dynamical}
\end{figure*}

As an example, we show an M67 model in which we strip mass from a 1.3 \Msolar~subgiant star at a rate of $10^{-5}$ \Msolar~yr$^{-1}$, stopping the mass
loss when the subgiant reaches 0.8, 0.9, 1.0, or 1.1 \Msolar. For an NGC 6791 model, we
strip mass from a 1.1 \Msolar~star, stopping mass loss when the subgiant
reaches 0.7, 0.8, or 0.9 \Msolar. Plots of the resulting MESA evolutionary tracks are
shown in Figure \ref{NGC6791_dynamical}. The high mass
loss rates produce models that are out of thermal equilibrium, causing a rapid drop in luminosity. When mass loss
is terminated, the models quickly return to equilibrium and resume evolution
along a subgiant/giant track. Due to their newly reduced masses,
these tracks are at lower temperature and luminosity than a normal cluster giant. A mass loss rate this large or larger would be required to strip enough envelope material to produce an SSG during a short-duration event like a dynamical encounter. However, lower mass loss rates between $10^{-6}$ and $10^{-8}$ \Msolar ~yr$^{-1}$ could also produce stars in the SSG region if the duration of the stripping event is longer. Lower rates produce a more gradual decline in luminosity and do not drive the star out of thermal equilibrium (similar to the mass transfer models in Fig. 3), but the models still move through the SSG region if they begin mass loss early enough on the subgiant branch to lose several tenths of a solar mass before beginning their ascent up the giant branch. These models indicate that mass loss of 0.3-0.4 \Msolar~on the subgiant branch can produce stars in the SSG domain for a wide range of mass loss rates. 

\subsection{Subgiant Collisions}
These MESA models indicate that if a subgiant loses significant envelope mass  it will move in to the SSG CMD region. We conjecture that one possible mechanism to remove this envelope mass would be a grazing dynamical encounter. In this encounter scenario, another star would have to pass close enough to a subgiant to tidally strip envelope material, but not close enough to lead to a merger. 

A similar mechanism has been proposed to explain the depletion of red giants near the Milky Way galactic center \citep{Dale2009}. In this scenario, encounters between RGB stars and black holes can eject the red giant core from the envelope. The core retains only a fraction of the envelope material, creating a giant with a significantly reduced envelope mass. They also find RGB-MS encounters capable of ejecting envelope material if the impact parameter is small enough. Similar models of encounters at the galactic center have found that encounters between RGB stars and MS stars, white dwarfs, or neutron stars can cause significant stripping of a giant envelope, though the amount of mass loss varies substantially between these studies. Depending on the specifics of the encounter and assumptions of the models, some conclude less than 10 \% of the envelope mass will be ejected \citep{Bailey1999}, and others find nearly the entire envelope may be lost \citep{Dale2009, Alexander1999}f.  At this stage of evolution, losing even a large fraction of the envelope does not prevent the giant from continuing its evolution up the giant branch, and it does not move into the SSG CMD region \citep{Dale2009}

Here we suggest similar encounters with subgiants or early giants may create SSGs. However, no existing studies focus specifically on subgiant encounters. New hydrodynamic simulations for subgiants would be necessary to advance this hypothesis, specifically determining the range of impact parameters that yield substantial mass loss while avoiding common envelope mergers as well as determining the possible orbital parameters for a post-encounter binary.

 \subsection{Frequency of Subgiant Dynamical Encounters}\label{sec:encounterrate}
 To explore the frequency of such dynamical encounters, we consider the case of the M67 SSGs. Using the encounter rates presented in \citet{Leigh2011}, we find the time between single-binary encounters to be $3.6 \times 10^8$ yrs. To get the encounter rate for subgiants, we can scale this rate by the fraction of stars in the cluster that are subgiants or early giants. In M67, we observe $\sim 30$ subgiants or early giants. Adopting the total number of stars in M67 to be $\sim 2000$ \citep{Geller2015} results in a subgiant fraction of $1.5\%$. Scaling our single-binary encounter rate by $1.5\%$, we find a time between single-binary encounters involving a subgiant to be $2.4 \times 10^{10}$ years. If we assume that all of these encounters lead to an SSG, and that the SSG is observable for its entire subgiant lifetime of $\sim 400$ Myr in M67, we find the Poisson probability of observing an SSG formed via a single-binary encounter in M67 to be $<2\%$. 
 
Using the same assumptions, binary-binary encounters also result in a $\sim2\%$ chance of observing an SSG. While encounters with triples may also play a role, the smaller number of triples makes binary-binary or single-binary encounters the dominant encounter types. More likely only a small fraction of encounters involving subgiants would strip the subgiant's envelope rather than leading to a merger, resulting in a very low probability of observing a dynamically formed SSG in M67. 

Dynamically formed SSGs may be more likely to be observed in larger clusters. We investigate this channel in more detail in the companion paper Geller et al. 2017b, including in globular cluster environments where the larger core densities and higher encounter rates may make this a more likely mechanism.



\subsection{Other Envelope Stripping Mechanisms}
We have proposed two mechanisms that create SSGs via mass loss from a subgiant: 1) mass transfer in a binary system and 2) tidal stripping of a subgiant's envelope during a dynamical encounter. Whether from mass transfer or envelope stripping, the essential finding in Sections 3 and 4 is that mass loss from a subgiant of several tenths of a solar mass successfully creates stars in the SSG CMD region. Our explorations suggest that neither Roche lobe overflow nor tidal stripping via dynamical encounters produce SSGs with high enough frequency to explain the observations. However, we have not fully explored all mechanisms of stellar mass loss, and there may be other ways for a subgiant to lose substantial envelope mass.  For example, if a binary system with a subgiant primary went through a period of common envelope evolution, ejecting some but not all of a giant's envelope material before the mass loss stabilized, the remaining system might resemble an SSG. \citet{Hurley2005} create an SSG in an N-body simulation of M67 using a similar mechanism. In their model, two stars merge in a common envelope event, ejecting 0.29 \Msolar~ masses of material in the process. The resulting star lies below the subgiant branch. As a second example, an SSG with a millisecond pulsar companion has been found in the globular cluster NGC 6397 \citep{Cohn2010, DAmico2001, Ferraro2003, Bogdanov2010}. The SSG is a giant that is extremely under-massive (0.22 \Msolar- 0.32 \Msolar) because it is being evaporated by the wind from its pulsar companion \citep{Ferraro2003}. While the N-body SSG is not in a binary and open clusters are unlikely to have millisecond pulsars, we encourage further exploration of more mass-loss hypotheses both in open cluster and other environments.

\section{Main Sequence Collisions}
It is also worth noting that dynamical encounters between main sequence stars are common, and can lead to the collision of two main sequence stars to form a single object. Such collision products are out of thermal equilibrium immediately after collision, and become much brighter than the main sequence due to the energy deposition in their envelopes during the encounter \citep{Sills1997, Sills2002}. As these products settle back into equilibrium, they contract and move back towards the main sequence. This occurs over a thermal timescale of a few Myr, but during this time such a collision product may be found in the SSG CMD domain \citep{Sills1997}. Due to the short duration of this phase compared to the single-binary and binary-binary encounter rate for M67 derived in Section~\ref{sec:encounterrate}, the Poisson probability of observing such a collision product is just a few percent. These encounter products are expected to be rapidly rotating single stars, not close binaries. Scattering experiments find that it is difficult for collision products to retain close binary companions (e.g. \citealt{Geller2013, Leigh2011}), and thus we do not consider this a likely explanation for the systems in our sample. While we do determine one of the SSGs to be a single star, it is not observed to be rotating rapidly, and thus is unlikely to be a recent collision product. Again, collisions are more common in denser globular clusters, and we explore the frequency of this formation mechanism in Geller et al. 2017b. 

\section{SSGs as Subgiants with Magnetically Inhibited Convection}
\subsection{Evidence for Magnetic Fields in the SSG Sample}
Five of the six stars show evidence that they possess strong surface magnetic fields: X-rays from a hot corona, H$\alpha$ emission from chromospheric activity, and starspots from areas of concentrated magnetic flux that inhibit convection and lower surface temperatures. 

 \citet{Belloni1998} determine 0.1-2.4 keV X-ray luminosities for the two SSGs
 in M67 of $7.3 \times 10^{30}$ \ergss~using ROSAT
 observations. \citet{vandenBerg2004} find a 0.3-7 keV luminosity of
 $1.3\times10^{31}$ \ergss for WOCS 13008 using Chandra-ACIS, while WOCS 15028
 is outside their field of view. \citet{vandenBerg2013} obtain Chandra
 observations of NGC 6791, detecting 3 of the 4 SSGs as 0.3-7 keV X-ray sources
 with luminosities of $1.27\times10^{31}$ \ergss (15561), $4.5\times10^{30}$
 \ergss (746), and $4.8\times10^{30}$ \ergss (3626). These X-ray luminosities
 are consistent with coronal emission due to magnetic activity. 
 
 All five of these SSG X-ray sources are also observed to be photometric variables with periods on the order of a few days. The lowest amplitude variable has $\Delta$V=0.02-0.09 depending on the variability survey (WOCS 170008; \citealt{Mochejska2002, Mochejska2005, Kaluzny2003, Bruntt2003, deMarchi2007}). The largest amplitude variable has $\Delta$V=0.26 (WOCS 130013; \citealt{deMarchi2007}). In all cases, the variability is attributed primarily to spot activity on the primary star, with perhaps ellipsoidal variations contributing in some cases. \citet{Milliman2016} present an overview of all known measurements of variability in the SSGs in NGC 6791. \citet{vandenBerg2002} present variability information for the 2 SSGs in M67. For 4 of these 5 stars, the variability is found to have periods close to (but not exactly at) the orbital period of the system. For the 5th system, 13008 in M67, they find variability, but do not have a time baseline long enough to determine the periodicity of the variability. 
 
 Finally, these 5 SSGs are also observed to have H$\alpha$ emission \citep{Milliman2016, vandenBerg1999, vandenBerg2013}, indicative of chromospheric activity. 

These 5 SSGs are the 5 binaries in our sample, all with orbital periods between 2.8 and 18.4 days. Taken together, these observations indicate these 5 SSGs are magnetically active binaries similar to RS CVn systems.
  
The 6th SSG (WOCS 131020) shows no evidence of X-ray emission, H$\alpha$
emission, or photometric variability. It is also the only one of the 6 stars
that is not observed to be a binary. Because of this lack of evidence for
magnetic activity in WOCS 131020, we suggest that the magnetic field
hypothesis is not well suited to explaining its
existence. \citet{Milliman2016} conclude that while it is unlikely all 4 SSGs
in NGC 6791 are field contaminants, it is possible that at least one of them
is. Perhaps 131020 is this field contaminant. It is also possible that WOCS
131020 is formed via a different formation channel that does not yield binarity, rapid rotation, and magnetic activity.

\subsection{The Impact of Spots on Stellar SEDs}
\citet{Mathieu2003} determine
spectroscopic temperatures a few hundred K warmer than our best-fit SED
temperatures in the M67 SSGs. One possible explanation is that the SSGs are known photometric variables, and spot activity on the
primary could skew the SED fits towards lower $T_\text{eff}$. Here we revisit our analysis of the SSG SEDs from Section 2 to analyze the impact of spots on the determination of radius, temperature, and luminosity. 

For example, we refit the
15028 SED to include a 3500 K spot while varying the spot covering
fraction. We perform this fit using the same approach detailed in Section 2.3, but add the spot covering fraction ($f_s$) as an additional free parameter. The 3500 K temperature is motivated by the known relation between
photospheric temperature and spot temperature  \citep{BerdyuginaReview}. A 4500-5000 K would
have a temperature contrast of $\sim$1000 K according to this relation. Here we model the SED as a combination of two SEDs weighted by the covering fraction ($f_s$), one representing the temperature of the unspotted photosphere, and the other a 3500 K SED representing the spotted photosphere. 

The new best-fit parameters for 15028 including a spot are T$_\text{eff}$=4750 K, R=3.1 \Rsolar, with a
3500 K spot (or spots) accounting for 40$\%$ of the surface area. Adding a spot to our model
provides a slightly better fit to the photometry and is consistent with the spectroscopically determined $T_\text{eff}$ of 4800 $\pm$ 150 K \citep{Mathieu2003}. Because the spectroscopic
temperature was measured in an optical spectral window, the flux from the star
would be dominated by the 4750 K surface rather than the 3500 K spot. 

Similarly, adding a spot to 13008 changes the best-fit parameters to 4750 K and 3.4 \Rsolar, closer to the 5000 K spectroscopic temperature \citep{Mathieu2003}.

\citet{Mathieu2003} also find a discrepancy between the radius of 15028 inferred from optical photometry (R=2.0 \Rsolar) and the radius inferred from geometry assuming a tidally synchronized rotation rate (R=4.0 \Rsolar). The best-fit radius assuming a spot is 3.1 \Rsolar, a substantially larger photometric radius than determined in \citet{Mathieu2003}. While this does not fully explain the discrepancy between the photometric and geometric determinations of the radius, it does bring the two radius measurements closer together. \citet{Mathieu2003} also observe a flux ratio between 15028's secondary and primary of 0.35, much higher than the expected ratio of 0.11 given the spectroscopic temperatures and geometric radii of the primary and secondary.  Our SED temperature and radius imply an expected V-band luminosity ratio of $\sim0.2$, closer to the observations but still lower than observed. 


Spot covering fractions for RS CVn have been measured using various methods (e.g. TiO band strength; \citealt{ONeal1996, ONeal1998, ONeal2004}, Doppler imaging; \citealt{Hackman2012}))
These measurements find covering fractions around 30-40$\%$ and sometimes up to 50$\%$ \citep{ONeal1996, ONeal1998, ONeal2004}, in line with the SED best-fits to the M67 SSGs. 

We conclude that if SSGs have a substantial spot covering fraction, the best-fit temperatures from our SED fits in Section~\ref{sec:SEDfits} may be too cool by a few hundred K and our best-fit radii may be too small by a few tenths of a solar radius. We encourage future observational efforts to determine the spot sizes and temperatures in order to better correct for this effect. 


\subsection{Modeling Inhibited Convection}
 While the interaction between magnetic fields and
convection remains unclear, theory suggests that a magnetic field may act to
reduce the efficiency of convective energy transport in stars with sufficiently large field strengths \citep{Stein1992}.
Observational evidence suggests the presence of magnetic fields in M-dwarfs
and solar-type binaries can create stars with  temperatures and radii that deviate from model
predictions. Radius determinations of eclipsing low-mass main-sequence stars (M $\lesssim 1.0$ \Msolar) are inflated by $5-10\%$ from model predictions \citep{Torres2002, Torres2006, Chabrier2005, Morales2008} and appear redder and cooler than typical low mass stars by a few percent \citep{Hawley1996}. Similar to our sample of SSGs, these stars display evidence of strong magnetic fields including X-ray and H$\alpha$ emission. Many have modeled these observations as an effect of inhibited convection due to the presence of magnetic fields\citep{Chabrier2007, Clausen2009, Feiden2013}. We suggest a similar effect may be at work in the SSGs.

Fully modeling the effect of magnetic fields on stellar evolution requires a 3D magnetic stellar evolution
code and is not currently technically feasible.  Instead, two main approaches have been used to produce 1D models of magnetically active stars. One approach is to introduce magnetic perturbations to the stellar structure equations, equation of state, and mixing-length theory of convection (e.g. \citealt{Feiden2012, Lydon1995}).  The other has been to reduce the mixing length coefficient ($\alpha_\text{MLT}$; \citealt{Chabrier2007}). The argument for using a reduced mixing length has been laid out in several papers \citep{Chabrier2007, Feiden2013}. In brief, the argument assumes that the star possesses a turbulent dynamo that sources energy directly from convective motions. The moving plasma in a stellar convective region induces a Lorentz force that preferentially opposes the movement of fluid across magnetic field lines. Thus the motion of a convective bubble will be slowed as some of its kinetic energy is diverted into the local magnetic field. For convective motions to be significantly slowed the local Alven velocity must approach the convective velocity. Given typical solar values, this requires internal magnetic field strengths of the order $10^4$ G, comparable to the equipartition field expected for rapid rotators. The result is a reduced heat flux transported by convection, which can be expressed in the framework of mixing length theory as a smaller characteristic convective length scale, or lowered mixing length coefficient $\alpha_\text{MLT}$.

\citet{Chabrier2007} simply alter $\alpha_\text{MLT}$ to fit the observed properties of magnetically active stars. They do not attempt to tie the value of $\alpha_\text{MLT}$ to a specific magnetic field strength or topology. In complementary work, \citet{Feiden2013} compare stellar models with reduced mixing length to models using a slightly different implementation of a turbulent dynamo, as well as to models using a magnetically modified Schwarzschild criterion. They also develop an expression relating the reduction in mixing length to a magnetic field strength. They find that the reduced mixing length models produce stellar structure nearly identical to their turbulent dynamo models, and that these models can reproduce the observed radius inflation among low mass main sequence stars using plausible internal magnetic field strengths and surface magnetic fields comparable to the observations.

Given the exploratory nature of this work, we elect to use the simple and easily implemented approach of \citet{Chabrier2007} of altering mixing length to match the observed temperatures and radii  of the sub-subgiants. However, the question of how magnetic fields impact stellar structure is far from settled. We consider this a useful first test, and expect future comparisons with sub-subgiant models using other approaches to implementing magnetic fields in stellar evolution codes will be necessary.

\subsubsection{MESA Models}

\begin{figure*}[htbp]
\subfigure[NGC 6791 R-T$_\text{eff}$ Diagram]{\includegraphics[width=.5\linewidth ]{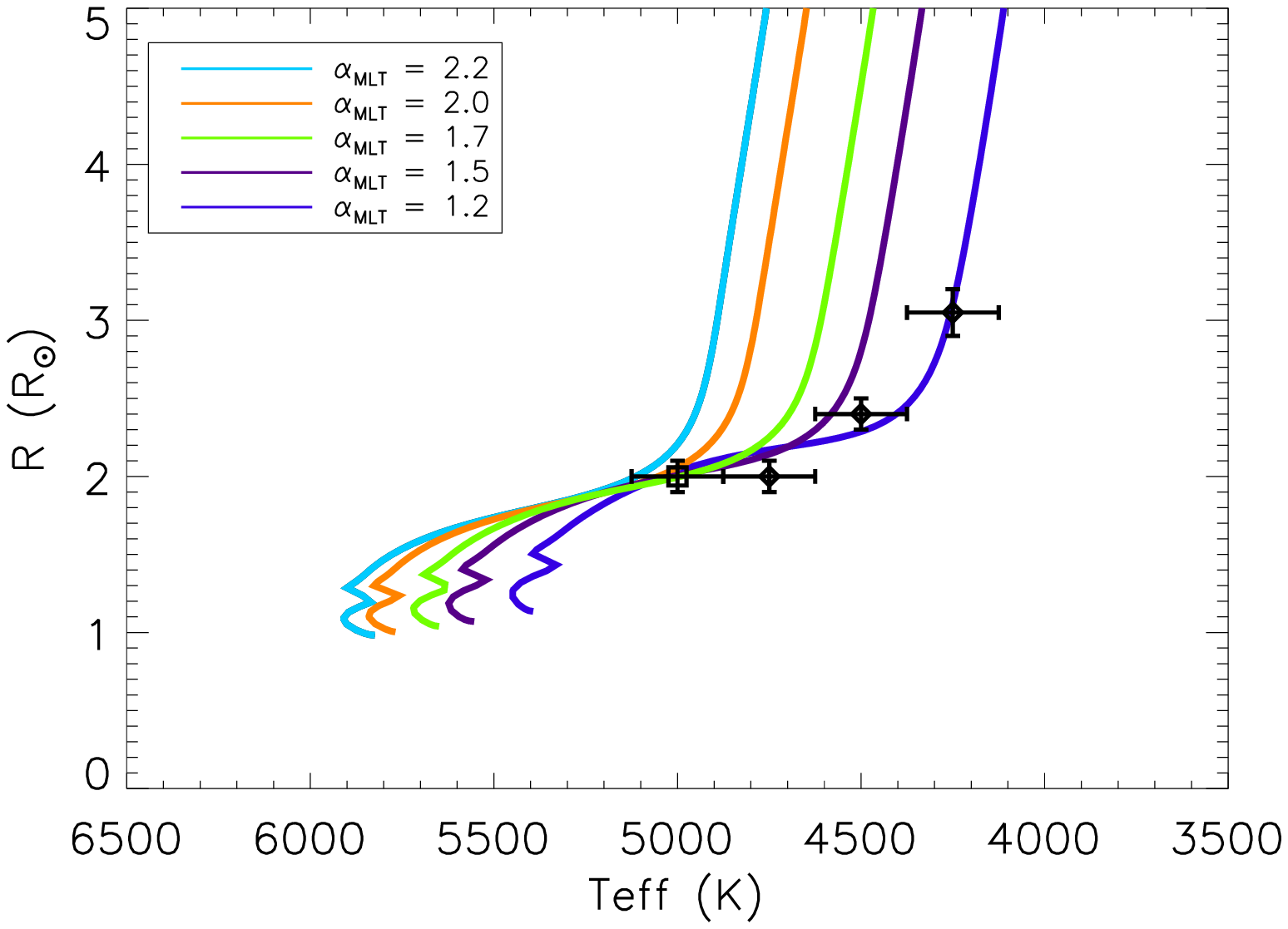}}
\subfigure[NGC 6791 HR Diagram]{\includegraphics[width=.5\linewidth ]{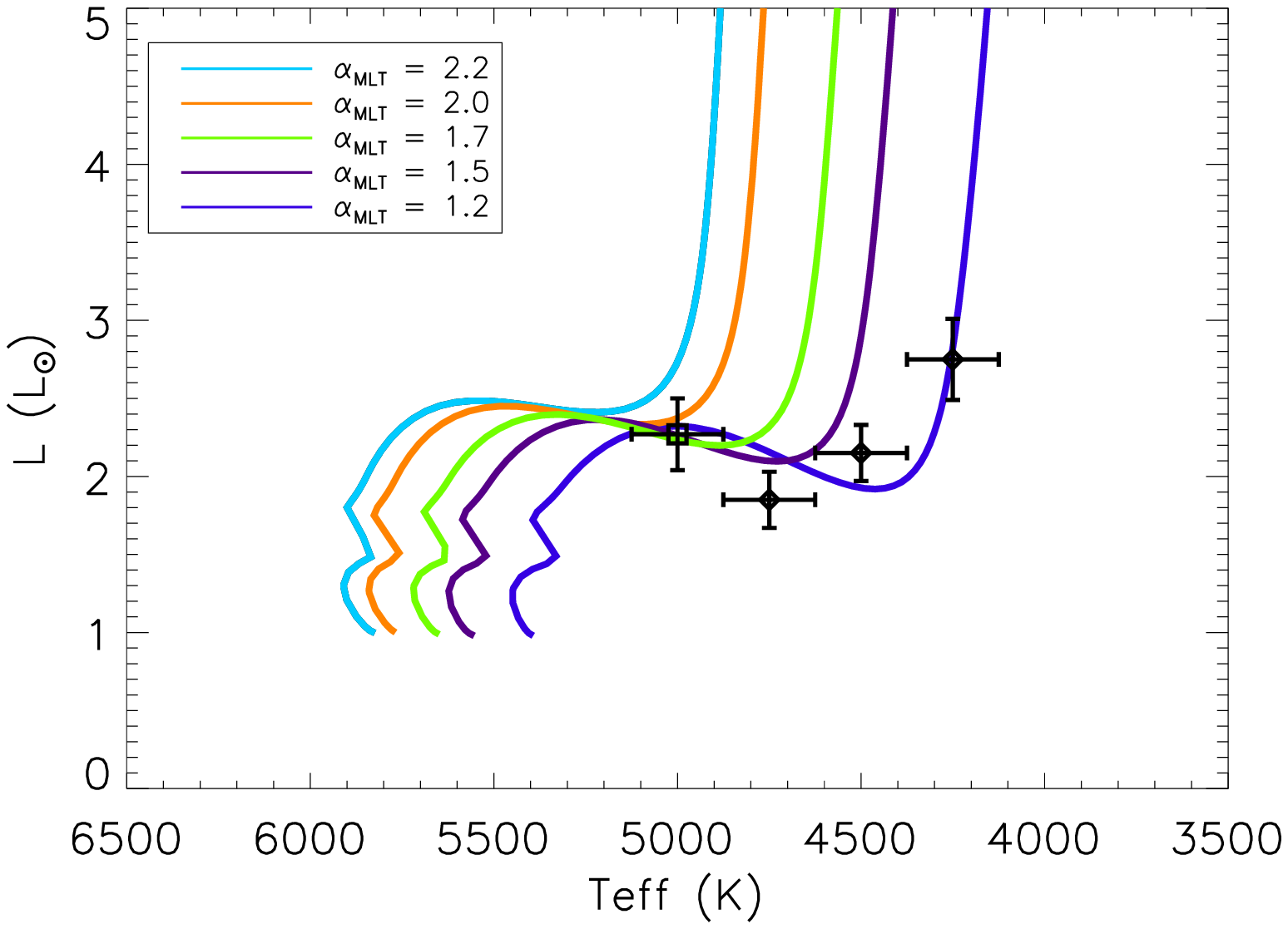}}
\subfigure[M67 R-T$_\text{eff}$ Diagram]{\includegraphics[width=.5\linewidth ]{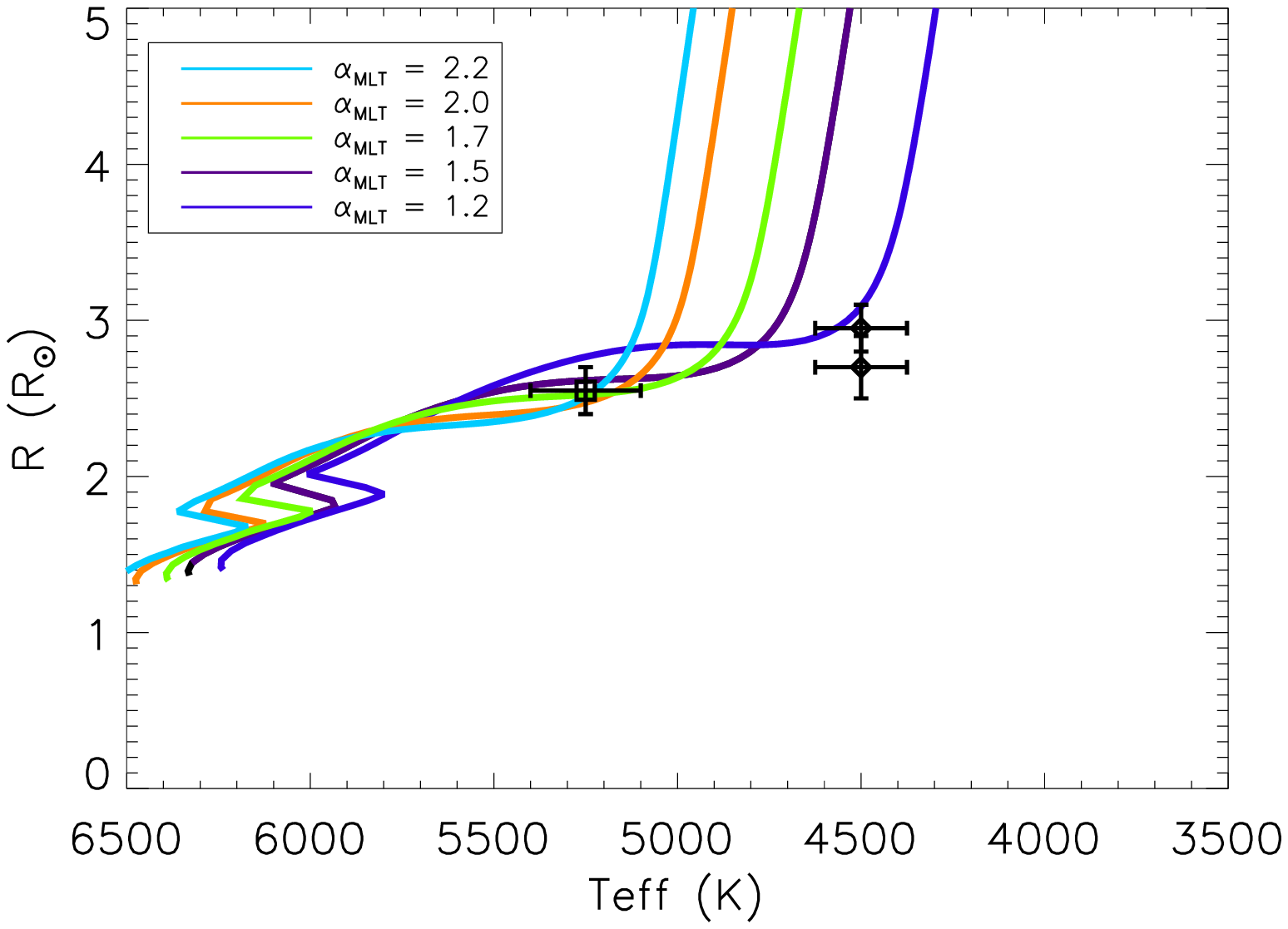}}
\subfigure[M67 HR Diagram]{\includegraphics[width=.5\linewidth ]{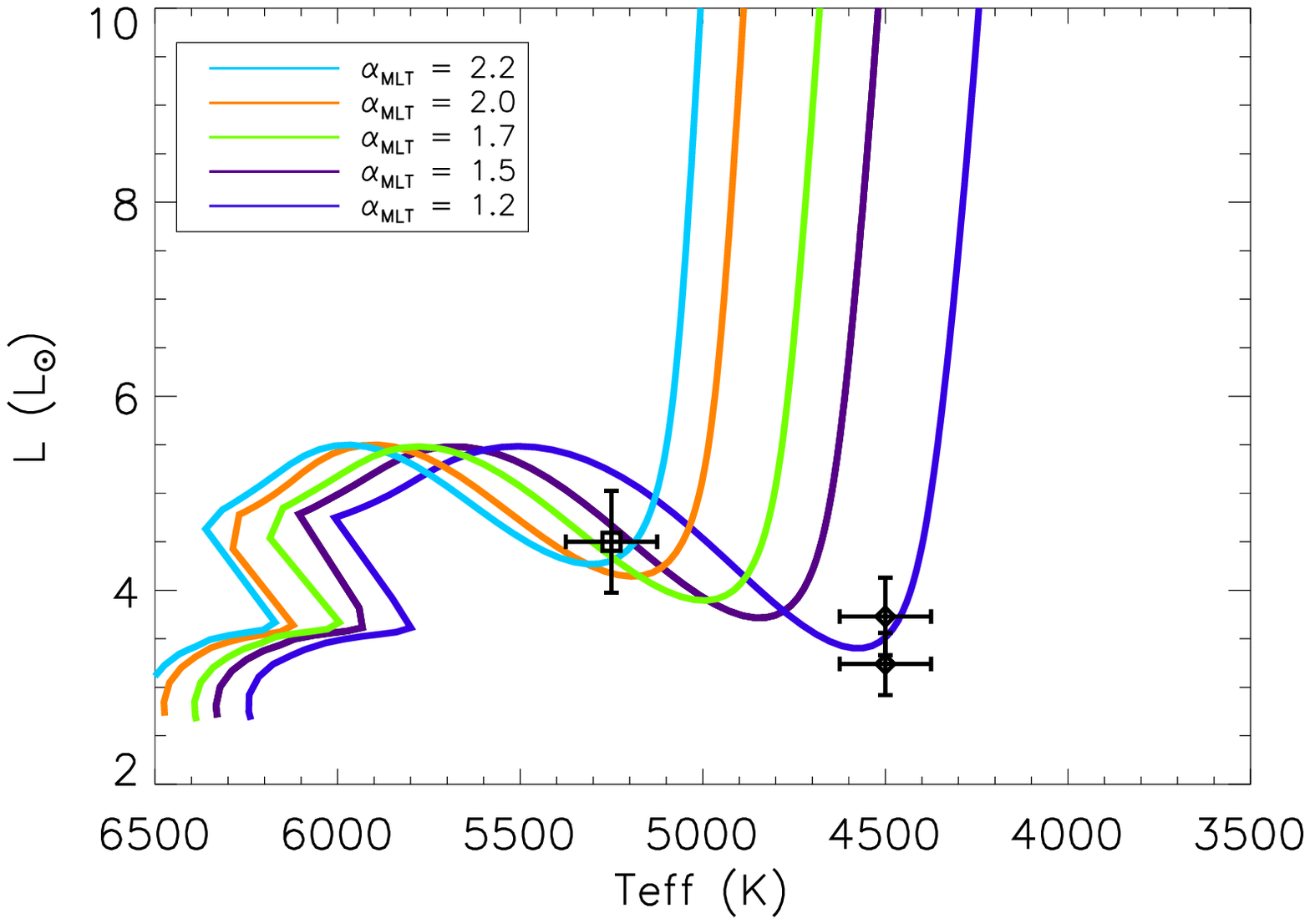}}
\caption{\textbf{(a)} an R-T$_\text{eff}$ plot with MESA models of the evolution of a 1.1 \Msolar~(the turn-off mass of NGC 6791) star given various mixing length coefficients ($\alpha_\text{MLT}$). The 3 magnetically active SSGs in NGC 6791 are plotted with diamonds. We do not include the 4th SSG, which does not show signs of magnetic activity. The location of a star at the base of the RGB in NGC 6791 (12270) is plotted for comparison with an open square.  \textbf{(b)} An HR diagram of the evolution of a 1.1 \Msolar~star with various mixing length coefficients ($\alpha_\text{MLT}$) from 2.2 to 1.2. Symbols are as in the diagram on the top left. \textbf{(c)} An R-T$_\text{eff}$ plot with MESA models of the evolution of a 1.3 \Msolar~(the turn-off mass of M67) star given various mixing length coefficients. M67 SSGs are plotted with diamonds. The location of a normal subgiant in M67 (10006) is plotted for comparison with an open square. \textbf{(d)} An HR diagram of the evolution of a 1.3 \Msolar~star with various mixing length coefficients. Symbols are as in the diagram on the bottom left. }\label{MLTmodels}
\end{figure*}

Following the approach of \citet{Chabrier2007},
we run a 1.3 \Msolar~model and a 1.1~\Msolar~model in MESA using different mixing length coefficients
in order to explore the impact this may have on the star's global
properties. MESA's standard mixing length coefficient is $\alpha=2.0$. Groups that model this reduced
convective efficiency in M-dwarf models found they required $\alpha$ to be
around 0.5 to reproduce the observed mass-radius relationship derived from
M-dwarf eclipsing binaries when assuming a uniform photospheric temperature. \citep{Feiden2013, Chabrier2007, Morales2010}. A larger mixing length of $\alpha=1$ was sufficient to reproduce observations when using a two temperature model for the photosphere to account for the effects of starspots to lower surface flux \citep{Chabrier2007, Morales2010}. 

In Figure~\ref{MLTmodels} we show the evolution of an SSG using various mixing
length parameters in a Hertzprung-Russell (HR) diagram and a R-T$_\text{eff}$ diagram. Also
plotted with diamonds are the SSG best-fit
radii, temperatures, and luminosity from the SED fitting assuming an unspotted surface (see Section~\ref{sec:SEDfits}). We compare to the unspotted SED temperature because the MESA models use a single temperature photosphere, and because the spot filling factors for the SSGs are still uncertain. We also plot for comparison the location of a normal cluster star located at the base of the RGB (open square). 

The models indicate that lowering mixing lengths creates cooler, larger stars at all points during the evolution of the star, but the luminosity remains unchanged at most stages of evolution regardless of mixing length. However, the altered mixing length does create lower luminosity stars near the end of the subgiant branch and the beginning of the RGB. At this stage in evolution, the expanding shell absorbs enough energy to lower the luminosity for a time. Lowering the mixing length parameter leads to a greater dip in the luminosity here, and this dip occurs at lower $T_\text{eff}$. The SSGs fall near this dip closest to the $\alpha=1.2$ track. 

\begin{figure*}[htbp]
\subfigure{\includegraphics[width=.5\linewidth ]{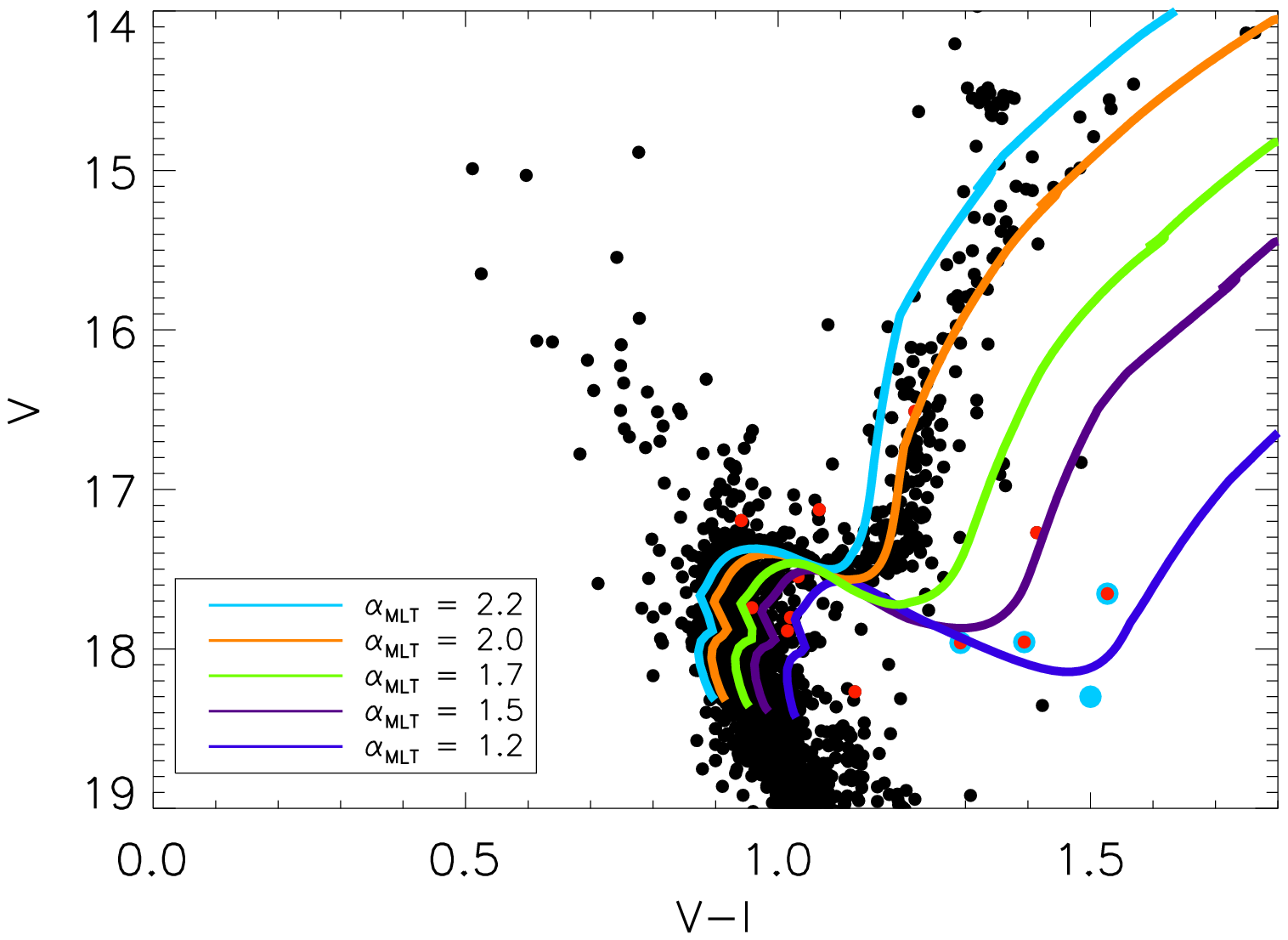}}
\subfigure{\includegraphics[width=.5\linewidth ]{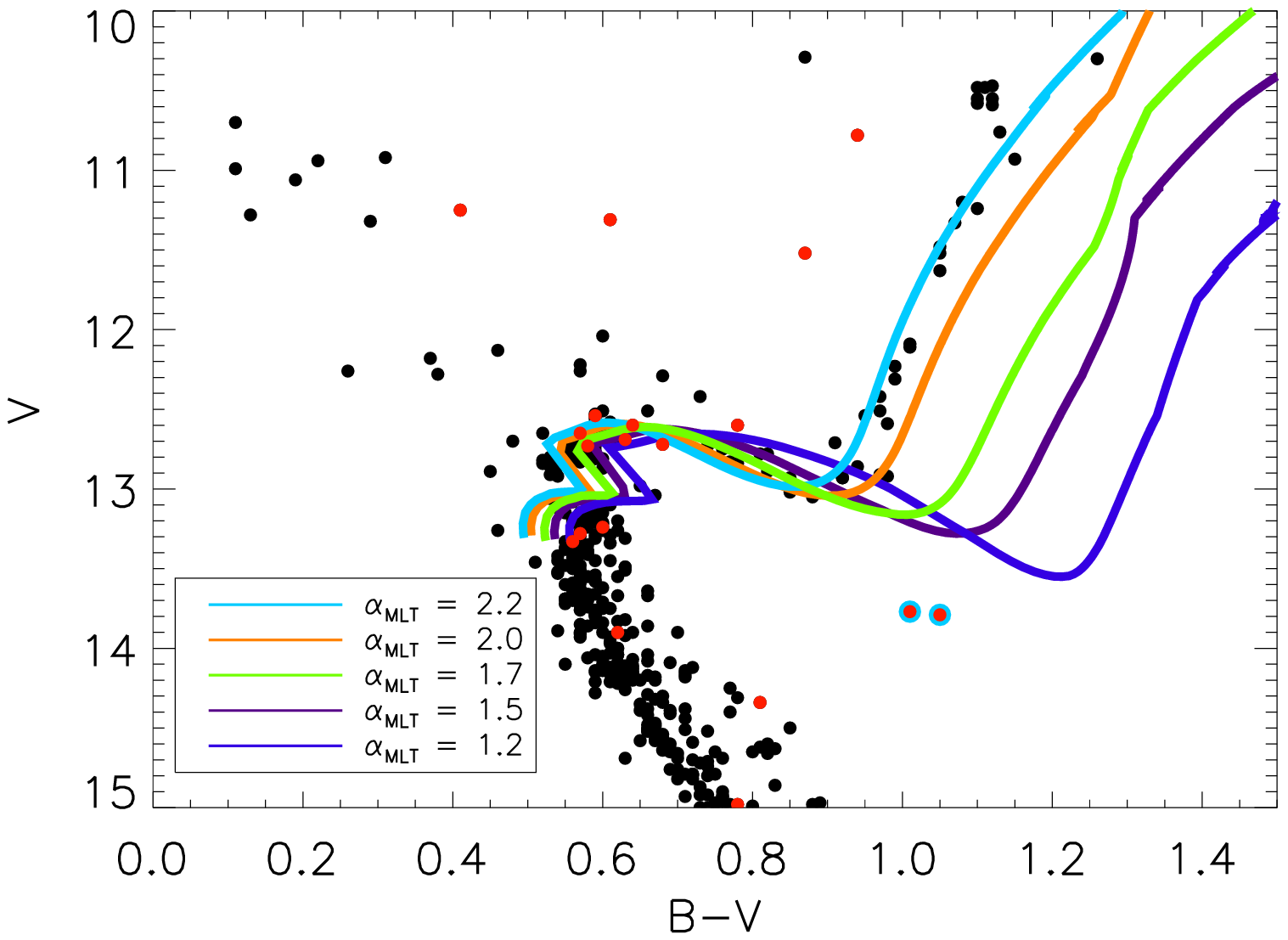}}

\caption{\textbf{(left)} A VI CMD of NGC 6791 showing all proper motion members (P$_\text{PM} \geq 95$) \citep{Platais2011}. X-ray sources from \citet{vandenBerg2013} are plotted in red and the 3D kinematic member SSGs are shown with larger light blue circles. Evolutionary tracks show the evolution of a 1.1 \Msolar~star using varying mixing length coefficients ($\alpha_\text{MLT}$) from 2.2-1.2. Colors indicate the different values from $\alpha_\text{MLT}$ as in Figure~\ref{MLTmodels}. \textbf{(right)} A BV CMD of M67 showing all 3D kinematic members \citep{Geller2015}. X-ray sources from \citet{Belloni1998} and \citet{vandenBerg2004} are shown in red and the two SSGs are shown with larger, light blue circles. Evolutionary tracks show the evolution of a 1.3 \Msolar~star, again using varying mixing length coefficients with the same values as in Figure~\ref{MLTmodels}. }\label{MLTCMD}
\end{figure*}

We also compare the CMD locations of SSGs to our models with lowered mixing length in Figure~\ref{MLTCMD}. The color transformation is done using the MESA colors module. This transformation assumes a uniform, unspotted surface temperature for the star. A more accurate treatment of the star would be to include both a lowered mixing length coefficient, and to assume a two (or more) temperature model for the surface flux that includes contributions from an unspotted photosphere and a spotted region. In fact, the temperature structure may be even more complex, with spots of different temperatures or hot plages surrounding the spots contributing to the emission. We therefore show the CMD to demonstrate the approximate region in which these lowered mixing length models would appear, but caution that the colors may not be accurate for highly spotted stars. 

While these models do move through the SSG region in a B-V or V-I CMD, we have less success producing the specific locations of the SSGs in M67 than in the HR diagram. These models predict stars that are redder, but brighter than the observed SSGs. We suggest this discrepancy may be due to our assumption of a single temperature photosphere in the MESA models. A better measurement of the spot temperatures and covering fraction of the SSGs and a stellar evolution code capable of modeling a spotted photosphere would provide more reliable color transformations. In NGC 6791, the models do better at reproducing the observed locations of the magnetically active SSGs. These stars fall between $\alpha=1.2$ and 1.5, similar to the results in the HR diagram. 

The CMD makes clear that changing mixing length becomes most noticeable near the end of the subgiant branch and through the RGB. While the tracks show cooler, more expanded stars on the main sequence and early subgiant branch, the spread in the tracks falls within the scatter of stars that fall on a normal isochrone, and therefore would not be noticed based on CMD position alone. We also expect that the such short-period magnetically active binaries are not observed all the way up the RGB, as they will evolve off the giant branch once they begin Roche lobe overflow. This model therefore predicts we should only observe SSGs in a small region just below or to the red of the lower RGB.

\subsection{Frequency of Formation}\label{frequency}


1) We start with the assumption that the binary fraction for systems with $P < 10^4$ days in both clusters is $25\%$. This is the average of binary fractions determined in other WOCS clusters of various ages: M35 ($24\%$, \citealt{Leiner2015}), NGC 6819 ($22\%$, \citealt{Milliman2014}) and NGC 188 ($29\%$, \citealt{Geller2012}). 

2) We assume that all binaries with periods less than 18 days (the longest period SSG in our sample) will produce magnetic fields on the subgiant branch that cause them to move through the SSG region. The tidal-circularization period for M67 is observed to be $\sim12$ days \citep{Meibom2005}, so this cutoff seems reasonable. 

3) We adopt the log-normal period distribution observed by \citet{Raghavan2010}. Integrating this distribution, we find that $11\%$ of binaries with $P < 10^4$ days have $P < 18$ days. Equivalently, we could say that $2.75 \%$ of all objects are binaries with $P < 18$ days. 

4) We count the number of objects observed in each cluster on the later half of the subgiant branch or the lower RGB. These are the areas predicted by MESA models to appear underluminous due to magnetic fields. In M67, where we have 3D kinematic memberships, we just count the number of stars in this region and find 20. In NGC 6791, we only have proper motion memberships. Here we take all members with P$_\text{PM} > 19 \%$ from \citet{Platais2011}. We correct for field contamination using the result of \citet{Tofflemire2014} that $73\%$ of the $P_\text{PM} > 19\% $ stars are confirmed as RV members. We find 100 members in our region of interest. 

5) Multiplying our number of stars by $2.75\%$, we expect to find 0.55 SSGs in M67 and 2.75 SSGs in NGC 6791 formed through this mechanism. 

6) We calculate the cumulative Poisson probability  to determine our odds of observing SSGs in M67 and NGC 6791.

We find that the chance of observing 1 or more magnetic SSGs in M67 is $42\%$. The chance of observing 1 or more magnetic SSGs NGC  6791 is $94\%$. The chance of observing 2 magnetic SSGs in M67 is $9\%$, while the chance of observing 4 magnetic SSGs in NGC 6791 is $15\%$. If we assume that only 3 of the 4 SSGs are magnetic since the 4th shows no signs of binarity or magnetic activity, we find a $22\%$ chance of observing 3 magnetic SSGs in NGC 6791.

We conclude that if this magnetic field mechanism can indeed create stars in the SSG CMD region, it is likely that several of the stars in our sample are created in this way. 


\section{Summary and Discussion}
Here we put forth three hypotheses of SSG formation: 1) Mass transfer in a binary system, 2) stripping of a subgiant's envelope, and 3) reduced luminosity due to magnetic fields that inhibit convection and produce large stellar spots. We demonstrate that stellar models for each of these methods evolve through the SSG domain.

Models of mass transfer in binaries containing subgiant stars can produce binary systems with combined light in the SSG CMD region. This requires binary systems with orbital periods around 1 day as they evolve along the subgiant branch, as longer period binaries begin RLO once they have started ascending the RGB and do not move through the SSG region. Additionally, the binary must have a small enough secondary that the secondary light does not push the combined light into a more populated CMD region (i.e. the blue straggler domain). Due to these restrictions on period and secondary mass, SSGs formed through mass transfer are expected to be rare, and we would not expect to see many, if any, in open clusters. However, with a larger sample of subgiant stars, e.g. in a massive globular cluster, we may observe SSGs formed in such a way. 

Furthermore, mass transfer models produce binaries with shorter periods than the observed orbital periods of the SSGs in M67 and NGC 6791. We test the tidally enhanced wind model of \citet{Tout1988}, and find that even with this elevated wind mass loss we cannot achieve the necessary mass loss rates to produce SSGs with the observed periods. 

MESA models in which several tenths of a solar mass of material is rapidly stripped from a subgiant's envelope can also produce stars in the SSG domain. We conjecture that this may happen during grazing dynamical encounters in which a star passes close enough to tidally strip material from a subgiant's envelope but avoids merging. Additional scattering experiments and hydrodynamic simulations are necessary to determine if this mechanism is viable, and if binaries with $\sim10$ day orbital periods are an expected end product of such an interaction. As an upper limit on formation rate, we assume that all single-binary or binary-binary encounters with subgiants lead to the formation of SSGs. Even with this very optimistic assumption, the expected rates of formation in open clusters are low enough that we would not expect to observe such stars in M67 or NGC 6791. SSGs may also form as the result of main-sequence collisions during dynamical encounters, but this too should be rare in open clusters. As with mass transfer, this may be a more relevant formation mechanism in massive globular clusters where encounter rates are higher and there are larger populations of subgiant stars, a possibility we investigate in another paper (Geller et al. 2017b, in preparation).

While envelope stripping during dynamical encounters and Roche Lobe overflow perhaps produce SSGs infrequently, clearly substantial mass loss from a subgiant star can create stars in the SSG CMD region. There may be other mass loss mechanisms that we have not explored that create SSGs more frequently, such as common envelope evolution. 

Stellar magnetic fields in-and-of
themselves may be sufficient to explain the anomalous luminosities and colors of
sub-subgiants. Five of our six SSGs show spot variability, X-rays, and H$\alpha$ emission indicative of surface
magnetic activity. Research in the field of
low-mass eclipsing binaries suggests magnetic fields can cause inflated
radii and lower effective temperatures. The SEDs of the SSGs suggest that
the stars do have lower temperatures than normal subgiants in the clusters. Simple 1D models
of stars with reduced mixing lengths can fairly well reproduce the observed
temperatures, radii, and luminosities of the systems. We have less success matching the models to the SSG CMD locations, which may be a consequence of assuming a single temperature photosphere when performing the color transformation. A stellar evolution code that allows for a multi-temperature photosphere is necessary to test these results and produce more accurate model colors, as are better measurements of spot temperatures and covering fractions (e.g. using TiO bands). 

 A calculation of the frequency with which magnetic fields should produce these stars yields an expectation of at least one such star in NGC 6791, and possibly one in M67. The formation frequencies indicate that several of the SSGs in our sample could likely have formed in this way, with a probability of a few percent that all five SSGs showing signs of magnetic activity could have been produced by this mechanism. The sixth SSG shows no signs of magnetic activity or binarity, and we conclude this mechanism is not likely to explain the origin of this system.
 
 Of course, these lowered-mixing length models are simplistic and do not include a physical treatment of the interaction between magnetic fields and convection. Without full 3D magnetic stellar evolution codes, fully implementing all the required physics is impossible, but several other stellar evolution codes use other approaches to model the effects of magnetic fields on evolution \citep{Feiden2014, Somers2015}. Comparing the results of these codes to our models would be a useful test of our approach. 

Finally, the discovery of more SSG SB2 binaries or eclipsing binaries in
order to infer masses would be an excellent test. While the
magnetic field hypothesis requires SSGs to be similar in mass to normal cluster
subgiants, mass-loss mechanisms such as mass transfer in a binary or envelope stripping require significant amounts of mass loss to produce SSGs. A sample of SSG systems in which the mass could be well determined would be a strong test of which hypothesis is best.

\acknowledgements
Thank you to the anonymous referee for helpful comments and suggestions. 

This research was supported by a Grant-In-Aid of Research from Sigma Xi, the Scientific Research Society and by NASA through grants HST-AR-13910 and HST- GO-13354.001-A from the Space Telescope Science Institute, which is operated by AURA, Inc., under NASA contract NAS 5-26555. This work was also funded by the National Science Foundation grant AST-0908082 to the University of Wisconsin-Madison. 
\bibliographystyle{mn2e}

\end{document}